\documentclass[preprint,review,letterpaper,12pt]{elsarticle}
\usepackage{amssymb}
\usepackage{amsthm}
\usepackage{amsmath}
\usepackage{bm}
\usepackage{algpseudocode}
\usepackage{algorithm}
\usepackage{multirow}
\usepackage{graphicx,subfigure}
\usepackage{caption}
\usepackage{graphics}
\usepackage{epstopdf} 
\usepackage{float}
\usepackage{eurosym}
\usepackage{anysize}
\usepackage{amsmath}
\usepackage{amsthm}
\usepackage{amsfonts}
\usepackage{fancyhdr} 
\usepackage{longtable}
\usepackage{amssymb}
\usepackage{color}
\usepackage{xcolor}
\usepackage{colortbl}
\usepackage{amssymb}
\usepackage{amsmath}
\usepackage{enumerate}
\usepackage{flushend}
\usepackage{lipsum}
\usepackage{float}
\usepackage{multicol}
\usepackage{verbatim}
\usepackage{natbib}
\usepackage{hyperref}
\usepackage{booktabs}       
\usepackage{tikz}
\usepackage{soul}  

\newcommand{\orcid}[1]{\href{https://orcid.org/#1}{\includegraphics[width=10pt]{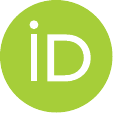}}}
\journal{arXiv}

\newfont{\tenbfit}{cmmib10}%
\newfont{\svnbfit}{cmmib8}%
\newfont{\tenbfsl}{cmbxti10}
\newfont{\mmit}{cmmi10}
\newfont{\smit}{cmmi9}
\newfont{\bfMit}{cmmi5}
\newfont{\tenbbb}{msbm10}%
\newfont{\svnbbb}{msbm8}%
\newfont{\tenssit}{cmssqi8 at 10pt}%
\newfont{\svnssit}{cmssqi8 at 7pt}%
\newfont{\gothic}{eufm10}%
\newfont{\sgothic}{eufm7}%
\newcommand{\Blj}{\mbox{$\Big[\kern-0.275em\Big[$}}
\newcommand{\Brj}{\mbox{$\Big]\kern-0.275em\Big]$}}

\begin{document}

\begin{frontmatter}

\title{Bayesian Calibration and Uncertainty Quantification of a Rate-dependent Cohesive Zone Model for Polymer Interfaces}
    
\author{Ponkrshnan Thiagarajan \orcid{0000-0003-3946-3902}} 
\author{Trisha Sain\corref{cor1}} 
\author{Susanta Ghosh \orcid{0000-0002-6262-4121} \corref{cor2}}
\address{Department of Mechanical Engineering-Engineering Mechanics, \\
Michigan Technological University\\
 Houghton, Michigan, USA}
\cortext[cor1]{Corresponding author, email:tsain@mtu.edu}
\cortext[cor2]{Corresponding author, email:susantag@mtu.edu}

\begin{abstract}
    In the present work, a  rate-dependent cohesive zone model for the fracture of polymeric interfaces is presented. Inverse calibration of parameters for such complex models through trial and error is computationally tedious due to the large number of parameters and the high computational cost associated. The obtained parameter values are often non-unique and the calibration inherits higher uncertainty when the available experimental data is limited. To alleviate these difficulties, a Bayesian calibration approach is used for the proposed rate-dependent cohesive zone model in this work. The proposed cohesive zone model accounts for both reversible elastic and irreversible rate-dependent separation sliding deformation at the interface. The viscous dissipation due to the irreversible opening at the interface is modeled using elastic-viscoplastic kinematics that incorporates the effects of strain rate. To quantify the uncertainty associated with the inverse parameter estimation, a modular Bayesian approach is employed to calibrate the unknown model parameters, accounting for the parameter uncertainty of the cohesive zone model. Further, to quantify the model uncertainties, such as incorrect assumptions or missing physics, a discrepancy function is introduced and it is approximated as a Gaussian process. The improvement in the model predictions following the introduction of a discrepancy function is demonstrated justifying the need for a discrepancy term. Finally, the overall uncertainty of the model is quantified in a predictive setting and the results are provided as confidence intervals. A sensitivity analysis is also performed to understand the effect of the variability of the inputs on the nature of the output.
\end{abstract}

\end{frontmatter}

\section{Introduction}\label{sec:intro}
Interfaces play a major role in dictating the overall mechanical performance of various composite structures and bi-material joints. Phenomena such as delamination in laminated composite systems \citep{doi:10.1177/0021998303034505}, failure of concrete dam-foundation joints \citep{barpi2010cohesive}, debonding of thin films from substrates \citep{LU20101679}, are a typical demonstration of interface failure. Such failures occur due to the local stress concentrations leading to separations and tangential sliding of the contacting surfaces across the interface. In the case of adhesively bonded components, the fact that the viscous or rate-dependent properties of the adhesives influence the global fracture response has been well documented in the recent literature  \cite{popelar1980dynamic, XU200315, du2000effects, SUN2009434}. In such cases,  the interfacial degradation depends on the rate of applied loading and the final response turns out to be rate-dependent as well.\par
In the case of polymer composite materials, crack initiation and propagation along the interfaces have been shown to be rate dependent \cite{smiley1987_1, smiley1987_2, kusaka1998}. It has been suggested that the bulk polymer viscous properties in general influence the global fracture response for the composites \cite{hui1992fracture}. To incorporate this rate-dependent behavior in the domain of computational modeling, various phenomenological cohesive laws have been proposed \cite{liechti2001, giambanco2006, marzi2009}. Earlier work by \cite{knauss1993} proposed a rate-dependent crack propagation model for craze-like fracture in polymers and failure of a joint bonded with a thin adhesive layer. In \cite{makhecha2009}, rate-dependent traction-separation relations were developed to simulate the stick-slip fracture in an adhesively bonded aluminum double-cantilever beam (DCB) specimen. In another work, a rate-dependent interface model was formulated considering a viscoplastic framework with hardening/softening behavior for shear and tensile traction \cite{giambanco2006}. Motivated by the experimentally observed differences in the nature of the propagating crack surfaces depending on the test speed, a nonlinear viscoelastic Kelvin model was introduced to simulate the rate-dependent cohesive response between rubber and steel at different rates under mixed mode loading condition \cite{liechti2001}. The rate dependence in both the bulk material and the interface was also considered in a similar model proposed by \cite{landis2000}. In \cite{marzi2009} a bilinear traction-separation relation was used with rate-dependent parameters to model the failure of structural adhesive joints under mode I loading. In that study, the parameters for the cohesive law were directly determined from experiments. As reported in \cite{mohammed2016}, experiments on pressure-sensitive adhesives were dominated by the rate-dependent interfacial properties, rather than the bulk viscoelasticity. 
Hence, the general agreement in the literature asserts the existence of rate-dependent fracture response in polymer-based interfaces. It can also be concluded that the overall rate dependence can arise as a consequence of the bulk material's behavior, of the interface response itself, or due to both.\par
The commonly utilized mathematical approach to study the interface fracture considers cohesive zone modeling \cite{elices2002cohesive}. To account for the complex microscopic processes that give rise to the new traction-free surfaces, cohesive zone models practically rely on the description of the traction-separation relationships. Such descriptions are phenomenological- but could be related to atomistic or molecular mechanisms \cite{ghatak2000, rahul1999, spearot2004}. Incorporating cohesive zones to model the interfaces of different materials, several research groups have demonstrated the capability of cohesive zone model (CZM) to track the complex crack propagation path \cite{needleman1990, tvergaard1992, xu1994, camacho1996, yang1999, gao2004, wei2009, wei2008, Su_2004, wei2010}, which otherwise could only be seen via tedious experiments. In particular, as reported in \cite{Su_2004} an elastic-plastic kinematic description was introduced to describe the irreversible separation-sliding behavior at the interface. Assuming the two contacting bodies as rigid, a yield function-based approach was proposed to describe the traction-separation behavior for both normal and tangential directions. Many of the rate-dependent cohesive zone models were developed under the assumption that the rate dependence arises only due to dissipation at the interfaces \cite{XU20039,XU200315,CORIGLIANO2001547}. One of the approaches focused on developing phenomenological constitutive laws that represent the cohesive strength and fracture energy as a function of opening/sliding rate at the interface \cite{corigliano2003numerical, anvari2006simulation,rosa2012loading}. These cohesive zone models are computationally less expensive than the models that assume a viscoelastic material ahead of the crack tip. However, most of these models were developed for a particular material system and loading conditions \cite{may2015rate, corigliano2003numerical, corigliano2006numerical}, limiting their applications.  Another group of the study had considered viscoelastic material models to characterize the rate-dependent bond breakage at the interface \cite{XU20039, MUSTO2013126, NME:NME4885, giraldo2017efficient}. The third group adopted viscoplasticity to capture the inelastic sliding separation at the interface prior to failure \cite{LU2013173,CORIGLIANO2001547}.
The most important aspect of the cohesive interface model in the context of the present work is the uncertainty associated with the estimation of the large set of model parameters. The commonly used approach of inversely identifying the cohesive zone parameters through nonlinear least square fitting is computationally prohibitive and often inaccurate due to limited experimental data. The major roadblock, in this case, is that the cohesive zone parameters are inaccessible via macroscopic fracture experiments. The existing literature is also limited  and rudimentary in quantifying the uncertainty in the CZM parameter estimations and how this parameter uncertainty would propagate in the final response. 

Uncertainty quantification for physics-based mathematical models is being intensely investigated since it can provide measures of confidence in the model prediction. In particular, the problem of parameter estimation by inverse calibration has remained central to uncertainty quantification. 
Kennedy and O$\textsf{'}$Hagan have pioneered a Bayesian approach for the calibration of the unknown parameters in a computer model \cite{kennedy2001bayesian}. Their model has received tremendous attention as a new approach for inverse calibration and is commonly referred to as the (KOH) approach. 
In the KOH approach, the discrepancy between the  computational model and the experimental observations is modeled explicitly by a discrepancy function. The true physical process is represented as a sum of the computational model, the discrepancy function, and the observational error. The computer model and the discrepancy function are treated independently and their priors are assumed to be Gaussian processes. The observational errors are assumed to be  zero mean Gaussians independent of each other. The posterior distributions of unknown model parameters and the discrepancy function parameters are estimated simultaneously using a Bayesian approach. Once these posterior distributions are estimated, the true process can be predicted along with the uncertainties associated with the predictions. \\

The KOH approach has been extensively investigated and further extended by several studies. A statistical approach (following the KOH approach) to combine scant field observations with simulation data for calibrating the unknown parameters in the simulation model and performing uncertainty quantification was demonstrated in \cite{higdon2004combining}.  In another work, Higdon \emph{et. al.} \cite{higdon2008computer} extended the KOH framework for computer simulations with multidimensional output. To overcome the challenges of size and the multivariate nature of the data, dimensionality reduction was performed using basis representations. A hierarchical Gaussian process model to combine data from multiple experiments with varying accuracies based on the KOH framework was introduced in  \cite{qian2008bayesian}. This model made use of the more abundant but less accurate data along with the less abundant high-accuracy data to produce predictions closer to the high-accuracy experiments. Arend  \emph{et. al.}  \cite{arendt2012quantification,arendt2012improving} illustrated the problem of identifiability, \emph{i.e.} whether the effects of calibration parameters and discrepancy function are distinguishable from one other in the model updating formulation and proposed a method to improve identifiability. An approach to calibrate the discrepancy function across different experimental settings based on the KOH framework was proposed in \cite{maupin2020model}.  A decoupled approach was introduced in which the unknown parameters of the computer model are estimated independently and prior to the estimation of the discrepancy function. This modular approach was computationally more feasible and it improved identifiability.  Several other noteworthy works based on the KOH framework are reported in  \cite{goldstein2009reified,bayarri2007framework,gramacy2008bayesian,ling2014selection}.

\par 
   The aforementioned Bayesian frameworks were employed for uncertainty quantification of diverse physics-based models such as plasticity models \cite{stevens2016experiment,asaadi2017computational,ricciardi2020uncertainty}, viscoelastic models \cite{miles2015bayesian}, turbulence models \cite{edeling2014bayesian}, and thermal models  \cite{liu2008bayesian,higdon2008bayesian}. For example, Asaadi et al \cite{asaadi2017computational} introduced a Bayesian framework for material characterization, involving both model class selection and parameter inference, in plasticity models. The framework integrated the Bayes' rule, surrogate modeling, principal component analysis, and nested sampling techniques. These works in the literature clearly demonstrate the potential of the Bayesian approach in quantifying uncertainties and calibrating parameters to improve the physics-based computational model. 

\par
Based on the current state-of-the-art, the objective of the present work is to perform uncertainty quantification for a phenomenological rate-dependent cohesive zone model. The proposed CZM is specifically designed to model the fracture response of the polymeric interfaces, based on an elastic-viscoplastic kinematical description. To enhance the robustness and accuracy of the proposed CZM prediction, uncertainty quantification of the model is further performed. To facilitate the inverse identification of the model parameters from limited experimental data the present study considers a Bayesian calibration approach. A sensitivity analysis is also performed to better understand the effects of inputs on the outputs of the CZM.  
\par
The rest of this paper is organized as follows:-In Sec.~\ref{sec:cons_model} the rate-dependent cohesive zone model for the polymeric interfaces is described, followed by an analytical implementation of the model in Sec.~\ref{sec:analytical imp}. Bayesian calibration,  Uncertainty quantification and Sensitivity analysis of the CZM are presented in Sec.~\ref{sec:bayes_cal} to Sec.~\ref{sec:sensitivity} followed by concluding remarks in Sec.~\ref{sec:conc}

\section{A Rate-dependent Phenomenological Cohesive Zone Model for Polymer Interfaces}\label{sec:cons_model}

\subsection{Kinematics}\label{subsec:kinematics}

\begin{figure}[htpb]
    \centering
    \includegraphics[width=0.8\textwidth]{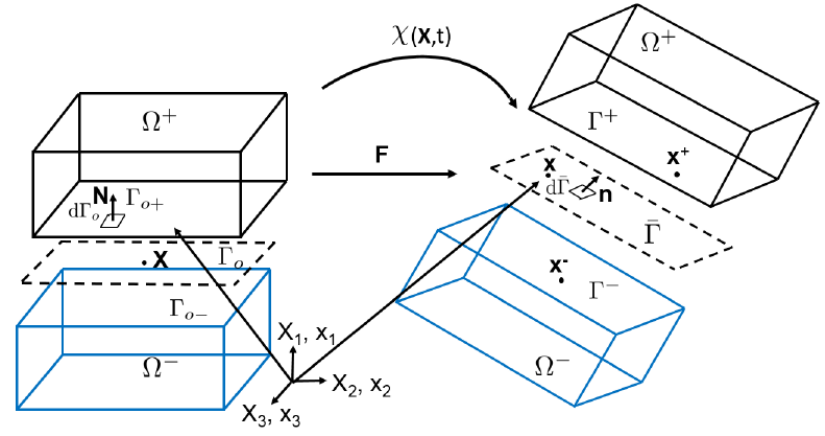}
    \caption{Schematic of an interface between two bodies $\Omega^{+}$ and $\Omega^{-}$.} \label{fig:interface}
    \end{figure}

\newcommand\myeq{\mathrel{\stackrel{\makebox[0pt]{\mbox{\normalfont\tiny def}}}{=}}}
In the present work, a rate-dependent traction-separation law has been proposed assuming a finite elastic-viscoplastic deformation of the polymer interfaces. In addition to that, a post-peak damage model has also been incorporated to model the degradation along the interfaces beyond post-yielding. The model has been proposed for a coupled normal and tangential (mixed mode) interfacial behavior along the interfaces. The phenomenological model stems from the work by
Su et al. \citep{Su_2004}, as mentioned in the introduction. The present work extends the model to incorporate a viscoplastic component to capture the rate-dependent behavior of the interfaces. Figure~\ref{fig:interface} represents the schematic of an interface undergoing finite opening and sliding.

Let us assume two bodies of polymeric materials $\Omega^{+}$ and $\Omega^{-}$ separated by an interface $\Gamma_o$ in the reference configuration as shown in Figure \ref{fig:interface}. The lower and upper surfaces are denoted as $\Gamma_{o-}$ and $\Gamma_{o+}$, respectively. In the reference configuration, the surfaces $\Gamma_{o-}$ and $\Gamma_{o+}$ are assumed to be identical to the reference interface $\Gamma_o$ where $\mathbf{X}_i$ represents the Cartesian material coordinates.

\begin{equation}\label{eq:ref_con}
\Gamma_{o} = \Gamma_{o-} = \Gamma_{o+}, \quad \Gamma_{o} = \Gamma_{o}(\mathbf{X}_i)
\end{equation}

In the current configuration, $\Gamma_{o+}$ and $\Gamma_{o-}$ become $\Gamma^{+}$ and $\Gamma^{-}$, respectively. A material point $\mathbf{X}$ initially on the interface $\Gamma_o$ in the reference configuration, is located on $\Gamma^\pm$ by the motion characterized by the displacement field $\mathbf{u}$ at time $t \; \in \; T$, where $T$ represents the time at which the deformation is applied.\par 

\begin{equation}\label{eq:current_con}
\mathbf{x}^\pm = \mathbf{X} + \mathbf{u}^\pm, \quad \Gamma_o \rightarrow \Gamma^\pm, \quad \forall t \; \in \; T
\end{equation}

where $\mathbf{x}^\pm$ denotes the material points on the upper and lower surfaces in the current configuration. Following the approach considered in \cite{ortiz1999finite} and \cite{xu2013elastic}, an interface $\Gamma$ is introduced in the current configuration to connect the strong discontinuities across the contacting surfaces consistently. The location of a material point $\mathbf{x}$ on the interface $\Gamma$, is defined by the uniquely invertible deformation map $\boldsymbol{\chi}$, as,

\begin{equation}\label{eq:mid_current_con}
\centering
\begin{split}
\mathbf{x} = \boldsymbol \chi (\mathbf{X},t), \quad \Gamma_o \rightarrow {\Gamma}, \quad \forall t \; \in \; T \\
\mathbf{x} = \frac{1}{2}(\mathbf{x}^{+}+\mathbf{x}^{-}), \quad \forall \mathbf{x}^\pm \; \in \; \Gamma^\pm
\end{split}
\end{equation}

Following which, the deformation gradient tensor $\mathbf{F}$ is defined as:

\begin{equation}\label{eq:def_grad}
\mathbf{F} = \frac{\partial \boldsymbol\chi(\mathbf{X},t)}{\partial \mathbf{X}},  \quad \forall \mathbf{X} \; \in \; \Gamma_o, \quad \forall t \; \in \; T
\end{equation} 

Hence, one can write that the interface $\Gamma_o$ with unit normal $\mathbf{N}$ is rotated and deformed to the interface $\Gamma$ having unit normal $\mathbf{n}$ in the current configuration by the following mapping:

\begin{equation}\label{eq:sur_map}
\mathbf{n} = (d\Gamma_o/d{\Gamma})\mathbf{F}.\mathbf{N}
\end{equation}
In a 3D representation, the cohesive zone is assumed to be a surface where displacement discontinuities occur as displacement jumps. Let us assume, $\bm{\delta}$ as the total displacement jumps across the cohesive interface. The displacement jump vector is defined by the following expression:

\begin{equation}\label{eq:dis_jump}
\bm{\delta} = \mathbf{x}^{+}-\mathbf{x}^{-},\quad \forall \mathbf{x}^\pm \; \in \; \Gamma^\pm, \quad \forall t \; \in \; T
\end{equation}
\subsection{Constitutive description for the traction-separation behavior}\label{subsec:cons_law}
Again, considering the framework in Su et. al \cite{Su_2004}, an additive  decomposition for the displacement jump vector is introduced as,
\begin{equation}\label{eq:disp_add}
\boldsymbol\delta = \boldsymbol{\delta}^e + \boldsymbol{\delta}^p
\end{equation}
where, $\bm{\delta}^e$ stands for the elastic displacement jump and $\bm{\delta}^p$ is the plastic, irreversible component of the same. To account for the rate-dependent inelastic behavior of the interface, a viscoplastic constitutive framework combined with a hardening and damage behavior is considered in the present work. The hardening behavior of the cohesive surface partially accounts for the defect evolution along the interfaces. To account for the post-yield damage in the interface, a scalar damage model is also considered.\\
Assuming, $\phi$ as the free energy per unit surface area in the reference configuration, based on a purely mechanical deformation, $\phi$ can be expressed as:
\begin{equation}\label{eq:phi}
\phi = \hat{\phi}(\bm{\delta}^e,\kappa,D)
\end{equation}
where $\kappa$ is a hardening variable, often expressed in terms of equivalent plastic strain/displacement. $\kappa$ describes the evolution of the interface yield surface and $D$ is the scalar damage variable. The time derivative of the free energy function is then given by,
\begin{equation}
\dot{\phi}(\bm{\delta}^e,\kappa,D)=\frac{\partial \phi}{\partial \bm{\delta}^e}.\dot{\bm{\delta}}^e+\frac{\partial \phi}{\partial\kappa}\dot{\kappa}+\frac{\partial \phi}{\partial D}\dot{D}
\end{equation}\label{eqn:phiDot}
Further, following the thermodynamic consistency, the dissipation inequality can be written as,
\begin{equation}
\bm{t}.\dot{\boldsymbol\delta}-\dot{\phi}\geqslant0
\end{equation}
where $\bm{t}$ is the traction vector.
Using Eqn.\ref{eq:disp_add} and \ref{eqn:phiDot} in the dissipation inequality we get,
\begin{equation}
\left(\bm{t}-\frac{\partial \phi}{\partial \bm{\delta}^e}\right).\dot{\bm{\delta}}+\frac{\partial \phi}{\partial \bm{\delta^e}}\dot{\bm{\delta^p}}-\frac{\partial \phi}{\partial\kappa}\dot{\kappa}-\frac{\partial \phi}{\partial D}\dot{D}\geqslant 0
\end{equation}
In order to satisfy the inequality for any arbitrary displacement jump, we pose,
\begin{equation}
\left(\bm t-\frac{\partial \phi}{\partial \bm{\delta}^e}\right).\dot{\bm{\delta}}=0
\end{equation}
Hence, the elastic traction-separation law for the cohesive interface can be obtained as,
\begin{equation}\label{eq:tr}
\bm{t} =  \frac{\partial \phi}{\partial \bm{\delta}^e}                    
\end{equation}
and the dissipation becomes,
\begin{equation}
\frac{\partial \phi}{\partial \bm{\delta^e}}\dot{\bm{\delta^p}}-\frac{\partial \phi}{\partial\kappa}\dot{\kappa}-\frac{\partial \phi}{\partial D}\dot{D}\geqslant 0
\label{eqn:dissip}
\end{equation}
Following Eqn. \ref{eqn:dissip}, a quadratic form of the free energy function is chosen as,
\begin{equation}
\phi=\frac{1}{2}(1-D)\bm{\delta^e}.\bm K. \bm{\delta^e}+H\kappa^2
\end{equation}
where the coefficient $H>0$ represents the hardening modulus and the matrix $\bm K$ denotes the interface elastic stiffness tensor as given by,
\begin{equation}
\bm{K} = K_N \bm{n} \otimes \bm{n} + K_T (\bm{1} - \bm{n} \otimes \bm{n})
\end{equation}
with $K_N>0$ and $K_T>0$ are the normal and tangential elastic stiffness moduli respectively.
Following equation \ref{eq:tr} the local traction vector is given by,
\begin{equation}
\bm{t} = (1-D)\bm{K}\bm{\delta^e} = (1-D)\bm{K}(\bm{\delta} - \bm{\delta^p})
\end{equation}
The interface traction $\bm{t}$ can be decomposed into normal component $\bm{t}_N$ and tangential component $\bm{t}_T$ as,
\begin{equation}
\begin{split}
\bm{t} = \bm{t}_N + \bm{t}_T \\
\bm{t}_N \equiv (\bm{n} \otimes \bm{n})\bm{t} = (\bm{t}.\bm{n})\bm{n} \equiv t_N \bm{n}\\
\bm{t}_T \equiv (\bm{1} - \bm{n} \otimes \bm{n})\bm{t} = \bm{t} - t_N \bm{n}
\end{split}
\end{equation}
where $t_N$ stands for the magnitude of normal stress at the interface. The magnitude of the equivalent tangential stress can be further written as:
\begin{equation}
\tau \equiv \sqrt{\bm{t_T} . \bm{t_T}}
\end{equation}
Here, $\tau$ is denoted as effective tangential traction. It is important to note that the displacement jump vector $\bm{\delta}$ has two components $\bm{\delta}_N$ and $\bm{\delta}_T$ corresponding to normal and tangential cohesive opening respectively.\\
In the 2D stress plane, the elastic domain of the cohesive constitutive law is defined as the interior of the convex yield surfaces. Once the applied interfacial displacements exceed the yield criteria, the response is governed by the choice of the yield function and the plastic flow rule. For a coupled normal and tangential cohesive behavior, the yield function is chosen as,
\begin{equation}\label{eq:yield}
\phi_Y={\tau}+\mu\langle t_N \rangle-S_{yp}
\end{equation}
where, $\langle t_N \rangle=0.5*(t_N+|t_N|)$ and $S_{yp}$ is the current yield strength, and $\mu$ is the friction coefficient. The yield strength evolution is given by the hardening law as,
\begin{equation}
S_{yp}=S_0+H.\kappa
\end{equation}
where $S_0$ is the initial yield stress and $\kappa$ is the hardening variable.
\subsection{Viscoplastic interface behavior, hardening law, and the post-yield damage}
In order to define the evolution laws for the internal variables associated with the dissipative phenomena, we need to define the flow rules for the plastic displacement jump $\bm{\delta}^p$, hardening variable $\kappa$ and damage variable $D$. As mentioned earlier, to model the rate-dependent interface behavior, a visco-plastic flow rule is adopted to describe the inelastic displacement jump as,
\begin{equation}
\dot{\bm{\delta}^p} = \dot{\gamma}_{vp}\bm{m}_{flow}
\end{equation}
with the plastic flow direction given by,
\begin{equation}
\bm{m}_{flow} = \frac{1}{\sqrt{1+\mu^2}} \left(\frac{\bm{t_T}}{\tau}+\mu\bm{n}\right)
\end{equation}
For a pure mode-I case, the 1st term in the bracket is led to zero and the flow direction is governed by the normal of the deformed interface. Similarly, the second term vanishes for pure shear loading, and the plastic flow direction is governed by tangential separation.\par
For the viscoplastic strain rate parameter $\gamma_{vp}$, a viscoplastic flow rule is considered as,
\begin{equation}\label{eq:visco_flow}
\dot{\gamma}_{vp} = \gamma_o \; exp\left(-\frac{Q}{k\theta}\left[1-\frac{\tau+\mu\langle t_N \rangle}{S_{yp}}\right]^{1/m}\right)
\end{equation}
where $\gamma_0$ is the reference plastic strain parameter, $Q$ is the activation energy, $m$ is the rate sensitivity parameter, $k$ is the Boltzmann constant and $\theta$ is the reference temperature.
As explained earlier, the rate-dependent behavior of the cohesive interfaces is critical to predicting the bi-material interface failure subjected to high rate loading. In polymeric materials, inelastic deformations are governed by thermally activated motions of macromolecules. Therefore, following the approach taken by Richeton et al. \cite{richeton2005formulation,richeton2006influence} and Ames et al. \cite{anand2009thermo, ames2009thermo}, a thermally activated relation is chosen to calculate the inelastic deformation rate as given by Eqn.\ref{eq:visco_flow}. It is to note that a considerably large number of visco-plastic models are found in the literature that accounts for plastic flow as a thermally activated process incorporating the temperature, strain, and the strain rate effects \cite{eyring1936viscosity, argon1973theory, mulliken2006mechanics}. Most of these models predict reasonably well the variation of the plastic strength as a function of temperature and strain rate within a limited range. However, it is seen that these models do not account for the sudden increase in yield stress at extremely high strain rates \cite{richeton2005formulation}. The flow rule in Eqn. \ref{eq:visco_flow} is motivated by the approach taken by Richeton et al. \cite{richeton2006influence, richeton2005formulation}. Their model is developed based on the ``co-operative'' model of Fotheringham and Cherry \cite{fotheringham1978role, fotheringham1976comment} which assumes that the flow in the polymer is allowed when several polymer chain segments are moving in a `co-operative' manner. A similar flow rule has also been adopted by Ames et al. \cite{ames2009thermo}. They have demonstrated that  such a viscoplastic model can predict the yield strength variation over a wide range of temperature and strain rates for amorphous polymers. In the present study, the assumption is that such a flow rule is adequate to predict the rate-dependent yield behavior of a thin layer of polymer adhesives as well.\par 
An evolution equation is further defined to describe the hardening variable $\kappa$, as
\begin{equation}
\dot{\kappa}=\dot{\gamma}_{vp}\frac{\partial \phi_Y}{\partial S_{yp}}
\end{equation}
In order to model the damage initiation and progression along the interface in the post-yield regime, the damage is assumed to be uncoupled from the plastic deformation. A simple damage rule, based on the total effective displacement is used as,
\begin{equation}
D=\frac{\delta^f(|\delta|-\delta^0)}{|\delta|(\delta^f-\delta^0)} \; \mathrm{for} \; \delta^0<|\delta|\leqslant \delta^f
\end{equation}
where $\delta^0$ and $\delta^f$ are the effective displacement jump at the onset of damage and at the final failure of the interface, respectively; $|\delta|$ represents the effective displacement jump defined as $|\delta|=\sqrt{\bm{\delta}_N^2+\bm{\delta}_T^2}$.

\section{Numerical implementation of the CZM} 
\begin{figure}
    \centering
      \includegraphics[width=0.7\textwidth]{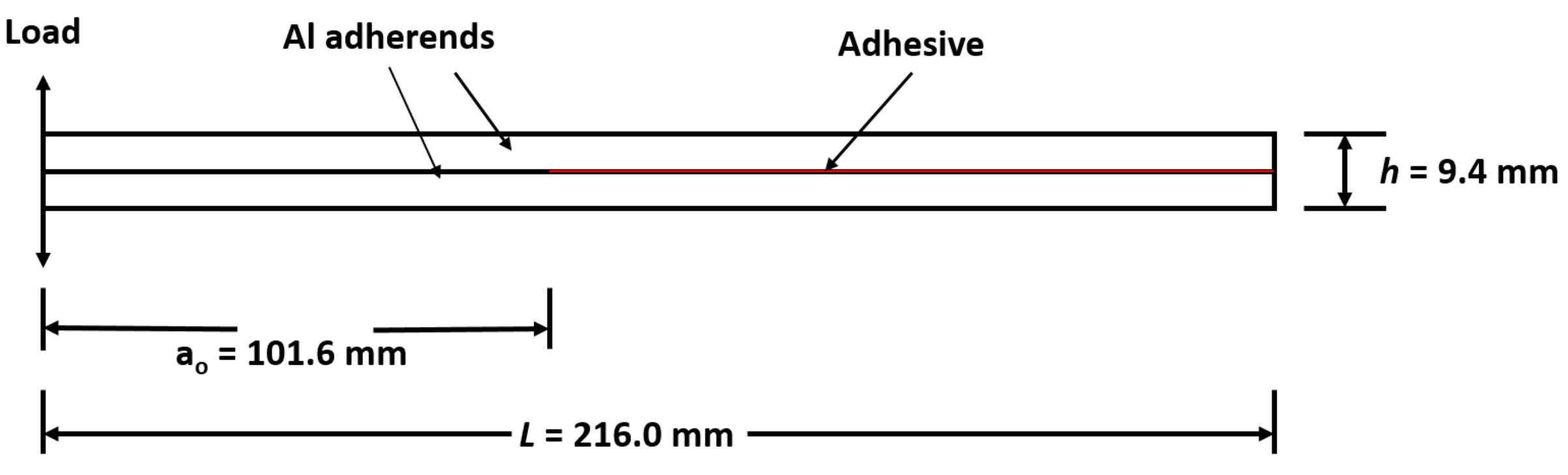}
    \caption{Geometry of the experimental DCB specimen as per \cite{XU200315};}
    \label{fig:Exp_DCB}
\end{figure}

To predict the interfacial failure of structural components, an analytical surrogate model is built to predict the mode-I interface failure using the proposed CZM. This analytical surrogate model considers a mode-I fracture geometry consisting of a 2-D double cantilever beam (DCB) specimen with an initial notch as shown in Fig~\ref{fig:Exp_DCB}. The uncertainty in this analytical surrogate model is quantified and presented in the following section.  We assume that the beams in the DCB geometry are almost rigid ($\approx 1000$ times stiffer) compared to the interface and the entire deformation only happens across the interface. This assumption would help to derive the analytical surrogate model. Model predictions are performed for three different displacement rates 5.08, 50.8, and 508.0 mm/min, respectively.  Prior to the discussion of results, we explain the analytical implementation of the proposed model and the inverse identification of the CZ parameters in the following subsections.
\subsection{Analytical surrogate model for the proposed cohesive zone model:-}\label{sec:analytical imp}  
\begin{figure}[htpb]
    \centering
     \includegraphics[width=0.45\textwidth]{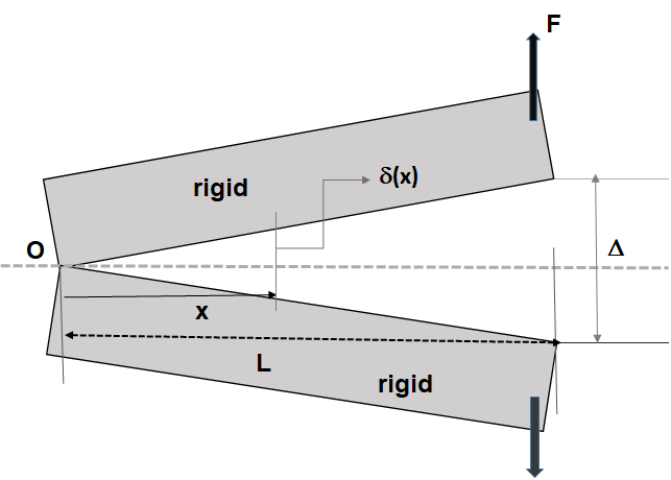}
              \caption{ Schematic of the rigid double cantilever beam under mode-I opening; }
        \label{fig:RDCB}
    \end{figure}

For the analytical implementation, the tangential sliding of the interface is ignored and the friction coefficient is assumed as $(\mu=1)$ to ensure the no-slip condition. Assuming the cantilevers as perfectly rigid, the (normal) opening along the interface line at any point $x$ from the pivot point can be estimated as, $\delta_N(x)=\frac{x}{L}\Delta$, where $L$ is the interface length, and $\Delta$ is the crack opening displacement along the applied load line as shown in Fig ~\ref{fig:RDCB}.
Balancing the moment exerted by the cohesive force generated due to the interface traction with the external moment due to the applied force $F$ about the pivot point ``O" we get, 
\begin{equation}
    B\int^{L}_0xt_N(x)dx=LF
\end{equation}
where $B$ is the specimen width and $t_N(x)$ denotes the (normal) traction at point $"x"$. The traction $t_N(x)$ is also a function of the interface opening at a distance $x$, as $t_N(\delta_N(x))$. It is also to be noted that the traction-separation law as described earlier is nonlinear in nature. Hence the closed-form integration for the moment balance equation is not trivial. Therefore, we numerically discretize the interface into a finite number of surface elements (in this case 1000) and calculate the traction distribution $(t_{N1},t_{N2}\ldots t_{N1000})$ for those elements in a discrete manner for a given displacement $\Delta$. These values are then used in the moment balance equation to calculate the applied external force $F$. \par
\subsection{Parameter estimation from experiments using the analytical surrogate model}
In the original experiments, rate-dependent debonding of a polyethylene-based adhesive had been studied using a double cantilever beam (DCB) set-up, similar to the geometry as shown in Fig~\ref{fig:Exp_DCB}.
The test specimen consists of Al601-T6 adherends bonded with a thermoplastic high-density polyethylene-based adhesive. Each adherend was 216 mm long, 4.70 mm thick, and 25 mm wide. The specimen contains an initial crack length of 101.6 mm, as shown in Figure \ref{fig:Exp_DCB}. During the experiments, displacement had been applied at the loading points along the direction indicated by arrows. The fracture behavior of the specimens was investigated and load-crack opening displacements were recorded at different cross-head displacement rates as 5.08, 50.8, and 508.0 mm/min, respectively. \\
The assumption of the ``rigid"ness of the bulk material in our analytical calculation considers the deformation only to happen across the interface. To incorporate the bulk deformation of the cantilever beams in the experiments, one needs to implement the CZM in a finite element framework which has not been considered in the present work. The focus of the present work is to consider the uncertainty associated with the model calibration in a rate-dependent phenomenological cohesive interface model in order to improve the accuracy and robustness of the model. Hence, we have only considered an analytical implementation of the proposed traction-separation law in a pure mode I condition and ignored the bulk material deformation.\\
In our analytical calculation, the thickness of the adhesive layer is considered zero.  To inversely determine the cohesive interface parameters a Bayesian calibration is performed. 
 The inverse calibration of parameters, from limited experiments always poses a non-uniqueness in the parameter estimation. The uncertainty associated with the model parameter estimation affects the accuracy of the model predictions for which experimental data is not available to verify.  In addition, determining the model parameters through the inverse trial-error process is computationally tedious, even for simple mode-I analytical calculation and the simulation time is a major bottleneck in the implementation of the model. Hence, a Bayesian estimation approach is proposed to calibrate the CZM parameters using limited experimental data.
\section{Bayesian Calibration} \label{sec:bayes_cal}
In general, computational models like the CZM take inputs $\bm{X}$ (strain rate and displacement in the CZM) to predict the quantities of interest $\bm{Y}$ (load at the specified displacement in the CZM). Where the inputs $\bm{X}$ to the model can be random or deterministic. Most computational models have additional parameters known as calibration parameters $\bm{\theta}$ (parameters provided in Table.~\ref{tab:prior_dis} for the CZM) that may or may not be obtained directly from experiments. Bayesian calibration is a powerful, mathematically founded and widely used method for identifying these unknown parameters of the computational model.  

\subsection{Methodology}
For the purpose of Bayesian calibration and uncertainty quantification, the experimental response ($\bm{Y}^{(e)}$) can be modeled following the Kennedy and O'Hagan approach \cite{kennedy2001bayesian} as:
\begin{equation}
    \bm{Y}^{(e)}  =  \bm{Y}^{(c)}(\bm{X},\bm{\theta}) + \bm{\delta}(\bm{X}) + \bm{\epsilon}
    \label{eq:koh}
\end{equation}
where, \\
$\bm{Y}^{(c)}(\bm{X},\bm{\theta})$ is the response from the computational model,\\
$\bm{\delta}(\bm{X})$ is the discrepancy between the model and the experimental response,\\
$\bm{\epsilon}$ is the uncertainty in the measurements of experimental response.\\
Owing to their computational efficiency, we use a modular approach \cite{maupin2020model} to calibrate the unknown parameters, $\bm{\theta}$, of the CZM  in this work. In this approach, the Bayesian calibration is performed separately and prior to the estimation of the discrepancy function.  Thus, the experimental response is written as,
\begin{equation}
    \bm{Y}^{(e)}  =  \bm{Y}^{(c)}(\bm{X},\bm{\theta}) +  \bm{\epsilon}
    \label{eq:calb}
\end{equation}
Where, the measurement error ($\bm{\epsilon}$) is modeled as a zero mean Gaussian,
$\bm{\epsilon} \sim N(0,\bm{\Sigma})$,  with covariance $\bm{\Sigma}$.

Therefore, the experimental response $\bm {Y}^{(e)}$ is Gaussian with mean $\bm{Y}^{(c)}(\bm{X},\bm{\theta})$ and covariance $\bm{\Sigma}$.
\begin{equation}
    \bm{Y}^{(e)} \sim N(\bm{Y}^{(c)}(\bm{X},\bm{\theta}),\bm{\Sigma})
    \label{eq:exp_dist}
\end{equation}

The unknown parameters $\bm{\theta}$ in \eqref{eq:exp_dist} can be estimated using a Bayesian approach. Given a prior distribution of the unknown model parameters $p(\bm{\theta})$ and a set of experimental observations $\bm{d}$, the posterior distribution of $\bm{\theta}$ can be estimated from Bayes theorem as follows:
\begin{equation}
    p(\bm{\theta} | \bm{d}) = \frac{p(\bm{d}|\bm{\theta}) p(\bm{\theta})}{p(\bm{d})} \label{eq:bayes}
\end{equation}
where, $p(\bm{d}|\bm{\theta})$ is the likelihood of observing the data $\bm{d}$ given the parameters $\bm{\theta}$. 
Given a set of $n$ independent experimental observations  $\bm{d} = \{ (\bm{x}_1,\bm{y}_1^{(e)}), (\bm{x}_2,\bm{y}_2^{(e)}),...(\bm{x}_n,\bm{y}_n^{(e)}) \}$, the likelihood function $p(\bm{d}|\bm{\theta})$ considering (\ref{eq:exp_dist}) can be written as,
\begin{align}
    p(\bm{d}|\bm{\theta}) &= \prod_{i=1}^n N(\bm{Y}^{(c)}(\bm{x}_i,\bm{\theta}),\bm{\Sigma})\\
    & = \prod_{i=1}^n \frac{1}{\sqrt{(2\pi)^{N_{out}} det(\bm{\Sigma})}} \exp \left(-\frac{1}{2}(\bm{y}_i^{(e)}-\bm{y}_i^{(c)})^T  \bm{\Sigma}^{-1} (\bm{y}_i^{(e)}-\bm{y}_i^{(c)})\right) \label{eq:likelihood}
\end{align}
With a prior distribution of calibration parameters, $\bm{\theta}$, the posterior distributions can be estimated from \eqref{eq:bayes}. The posterior distribution of the parameters provide the uncertainty in the model along with a point estimate for the model parameters $\bm{\theta}$.
\subsection{Prior distribution of parameters} \label{sec:priors}
The prior distribution of parameters represents our prior knowledge or assumptions about the parameter and they play a key role in Bayesian inference problems. The prior distribution has minor effects on the posterior when the experimental data is sufficiently large in number. Whereas, when the data is limited, as in the case of this work, the prior distribution of parameters plays a significant role in determining the posterior of parameters. Therefore it becomes crucial to make reasonable choices for the prior distribution.

The most common choices for the prior distribution are the non-informative uniform distribution and the informative Gaussian distribution. Uniform distribution for the parameters is assumed when no information about the parameter is available. Since the support of the uniform distribution is bounded, one has to be careful in choosing the bounds of the uniform distribution. On the other hand, if a value of the parameter is approximately known a Gaussian distribution can be assumed with the mean at that value. In this work, both the uniform and the Gaussian distributions are assumed for various parameters as explained below.

\subsubsection{Normal stiffness, $K_N$}
A Gaussian distribution is assumed for the parameter $K_N$. The parameters of the Gaussian distribution are evaluated from the experimental response of tough polyethylene in \cite{pandya2000measurement}. The slope of the traction separation curves for the polyethylene (PE2) is taken as the mean of the Gaussian and the standard deviation is assumed one-sixth of the mean. \footnote{$\sigma = \mu/6$ is assumed so that $3\sigma$ or $99.7\%$ of the samples are within  $[0.5\mu \;\; 1.5\mu]$.}

\subsubsection{Dispalcement at the onset of damage and final failure, $\delta_o, \delta_f$}
 From the experimental response of the DCB experiments, the peak load occurs at 10 mm COD. Therefore, the onset of damage is assumed to occur anywhere from 0 to 10 mm and the final failure is assumed to occur anywhere from 10 to 20 mm. Due to these assumptions, uniform prior distributions are assumed for these parameters with the corresponding range of values.

\subsubsection{Hardening modulus, $H$ and Activation energy, $Q$}
A Gaussian distribution is assumed for the hardening modulus and the activation energy. The parameters of the Gaussian distributions are taken from a similar constitutive model used for the polycarbonate materials in \cite{srivastava2010thermo}. The value of the hardening modulus and activation energy in \cite{srivastava2010thermo} is used as the mean of the Gaussian with a standard deviation of one-sixth and one-third of the mean respectively.

\subsubsection{Normal yield strength $S_o$}
The normal yield strength of polyethylene is taken from an online materials database for high-density polyethylene \cite{matweb}. The prior distribution is assumed Gaussian with value from the database as the mean and one-sixth of this value as the standard deviation.
\subsubsection{ Reference plastic strain $\gamma_o$ and Rate sensitivity parameter $m$}
Since $\gamma_o$ and $m$ are parameters of the phenomenological model introduced in this work for characterizing the viscoplastic interface behavior, no knowledge about these parameters exists. Therefore, a minimum number of trial and error is performed to get an approximate value (or order of magnitude) for these parameters. Using these values as the mean a Gaussian distribution is assumed for these parameters and a standard deviation of one-third and one-sixth of the mean is assumed for $\gamma_o$ and $m$ respectively.

With these assumptions and prior knowledge, the prior distribution of the eight unknown parameters of the cohesive zone model can be summarised as given in Table.~\ref{tab:prior_dis}. It is to be noted that the support of all the parameters is constrained to be non-negative. 
\begin{table}[htpb]
    \centering
    \begin{tabular}{lcc}
    \toprule
         Parameter &  \multicolumn{2}{c}{Prior}  \\ 
         &  Distribution &Parameters \\ \midrule
         Normal stiffness, $K_N$ (MPa/mm)  & Gaussian & $[240,40]$ \\ 
         Displacement at the onset of damage, $\delta^0$ (mm)   & Uniform & $[0,10]$ \\ 
         Displacement at final failure, $\delta^f$ (mm) & Uniform &$[10,20]$ \\ 
         Hardening  modulus, $H$ (MPa/mm)  & Gaussian & $[58,9.67]$ \\ 
         Normal yield strength, $S_o$ (MPa) & Gaussian & $[60.7,10.12]$ \\ 
         Reference plastic strain,  $\gamma_o$ (mm/s)  & Gaussian & [1e-6,0.33e-6] \\ 
         Activation energy, $Q$ (N-mm) &Gaussian & [1.5e-19,0.5e-19] \\ 
         Rate sensitivity parameter, $m$ & Gaussian & $[25,4.17]$ \\ 
         \bottomrule
    \end{tabular}
    \caption{Prior distribution of parameters used for Bayesian calibration}
    \label{tab:prior_dis}
\end{table}
\subsection{Results of the Bayesian calibration}
The analytical implementation of the CZM, described in  Sec.~\ref{sec:analytical imp}, that simulates the load-displacement curve of the DCB experiment is used for the calibration of unknown parameters. The CZM is integrated with the Bayesian inference module of UQLab \cite{UQdoc_13_114} in MATLAB for this purpose. The CZM acts as the forward computational model $\bm{Y}^{(c)}(\bm{X},\bm{\theta})$ in equation \eqref{eq:calb}, whose input ($\bm X$) is the displacement and the output $\bm{Y}^{(c)}$ is the load. There are eight unknown model parameters ($\bm{\theta}$) that are calibrated. 

In order to calibrate the unknown parameters, experimental results of the DCB experiments and their corresponding CZM predictions are obtained for three different strain rates. Specifically, 20 points from the load-displacement curve are considered from the experiments and the model predictions.  The posterior distributions of the model parameters are obtained from equation (\ref{eq:bayes}) by using the experimental data, the model prediction, and assuming prior distributions of parameters. The posterior distribution in Eqn (\ref{eq:bayes}) is intractable and thus it is approximated via a Markov Chain Monte Carlo sampling method. An Affine Invariant Ensemble Algorithm (AIES) is used to obtain the posterior distribution of the unknown parameters. 




With the priors for the unknown model parameters assumed as described in Sec.\ref{sec:priors}, the estimates for the posterior of parameters presented in Table \ref{tab:calb_res} are obtained using the Bayesian calibration approach.
\begin{table}[htpb]
    \centering
    \begin{tabular}{lcc}
    \toprule
         Parameter &  \multicolumn{2}{c}{Posterior}  \\ 
         &  Mean &Std \\ \midrule
         Normal stiffness, $K_N$ (MPa/mm)  & 326.81 & 0.1815 \\ 
         Displacement at the onset of damage, $\delta^0$ (mm)   & 5.83 &1.72e-3 \\ 
         Displacement at final failure, $\delta^f$ (mm) & 17.91 &0.0594 \\ 
         Hardening  modulus, $H$ (MPa/mm)  & 0.3376 & 0.0805 \\ 
         Normal yield strength, $S_o$ (MPa) & 78.87 & 11.87 \\ 
         Reference plastic strain,  $\gamma_o$ (mm/s)  & 3.7e-7 &5.75e-8 \\ 
         Activation energy, $Q$ (N-mm) &1.58e-19 & 3.02e-20 \\ 
         Rate sensitivity parameter, $m$ & 47.06 & 6.01 \\ 
         \bottomrule
    \end{tabular}
    \caption{Mean and standard deviations of the Gaussian priors and posteriors of Bayesian calibration}
    \label{tab:calb_res}
\end{table}

\begin{figure}[htpb]
    \centering
    \subfigure[5 mm/min strain rate]{\label{fig:bayes_a}\includegraphics[width = 0.48\linewidth]{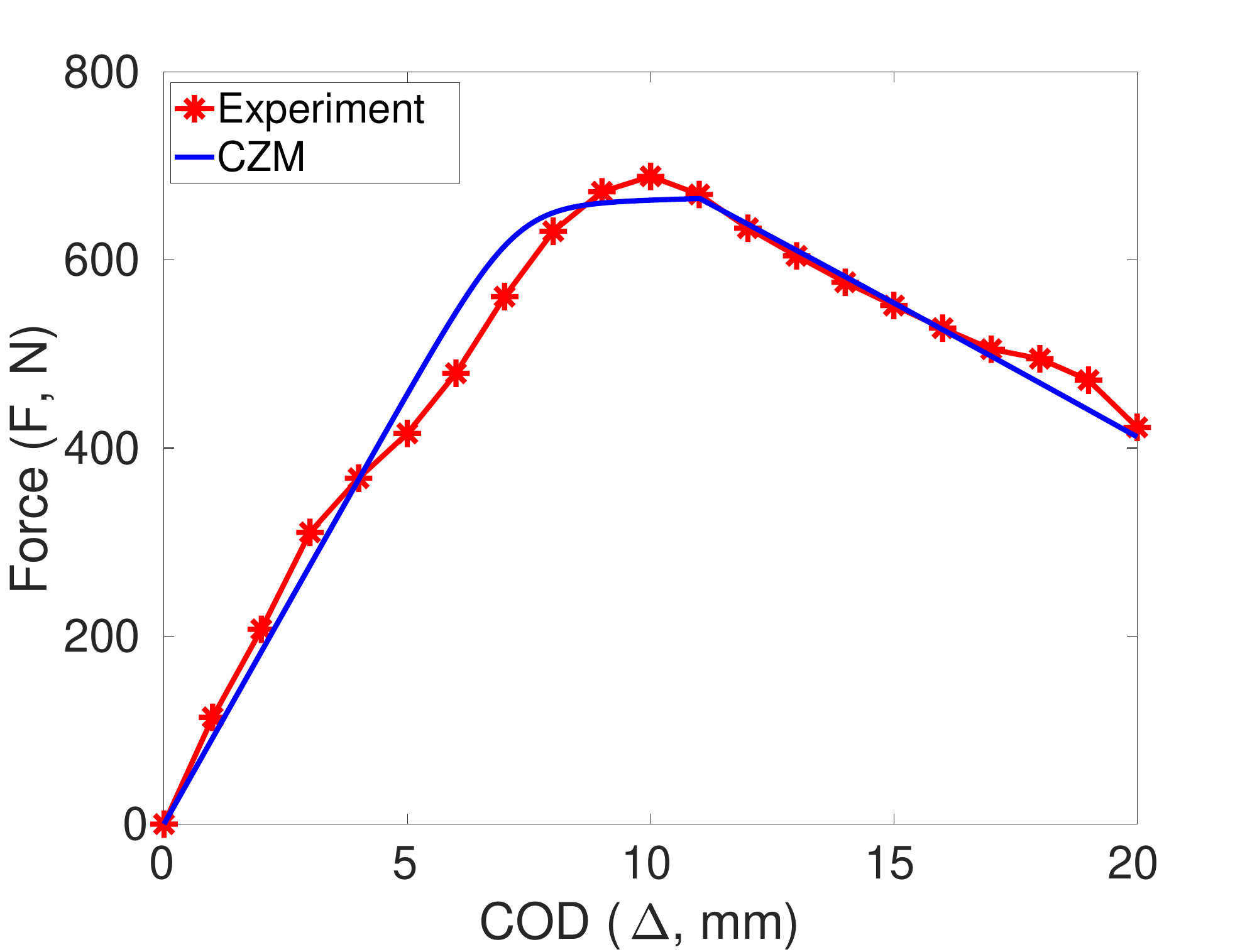}}
    \hfill
    \subfigure[50 mm/min strain rate]{\label{fig:bayes_b}\includegraphics[width = 0.48\linewidth]{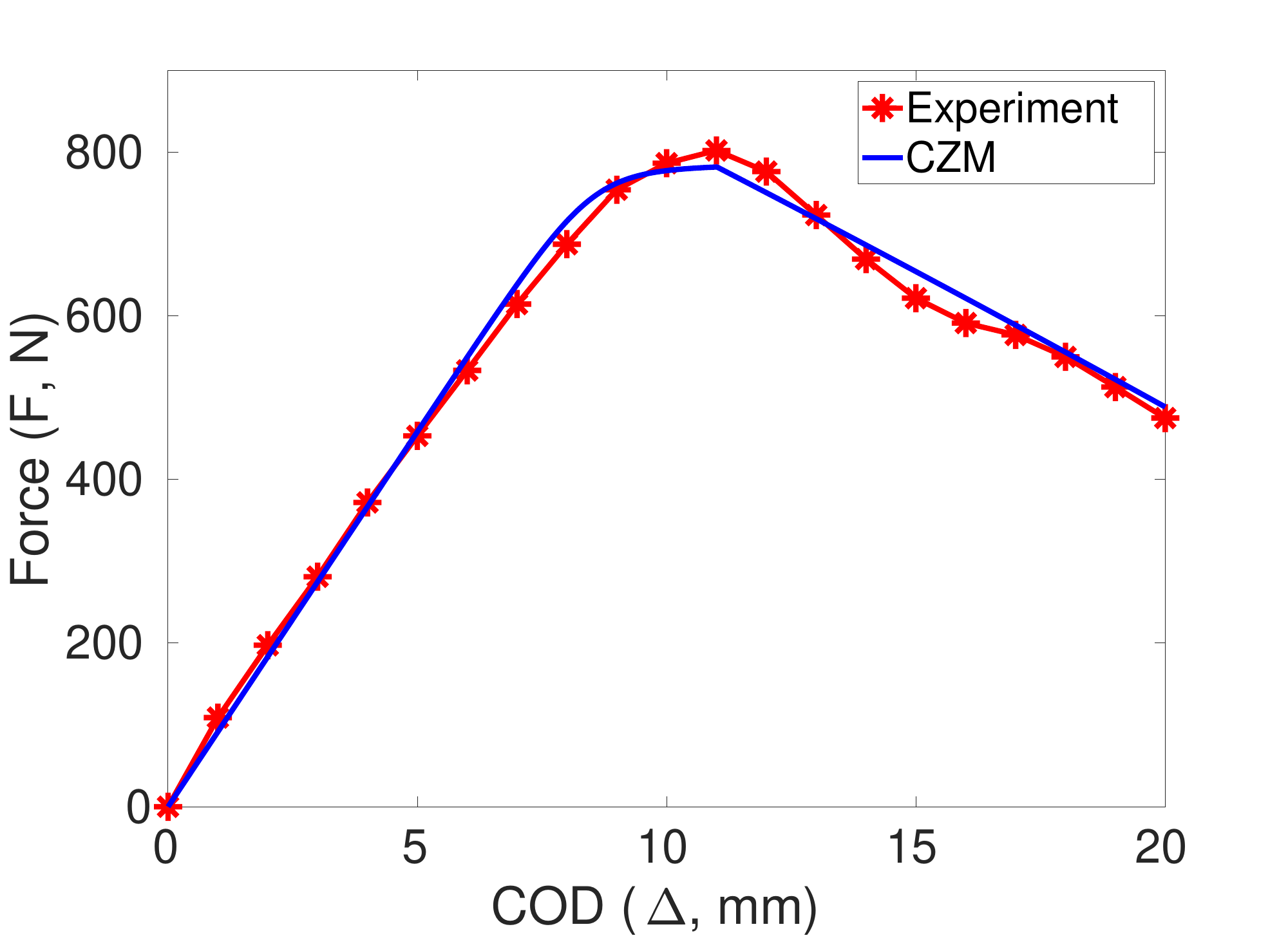}}
    \subfigure[500 mm/min strain rate]{\label{fig:bayes_c}\includegraphics[width = 0.48\linewidth]{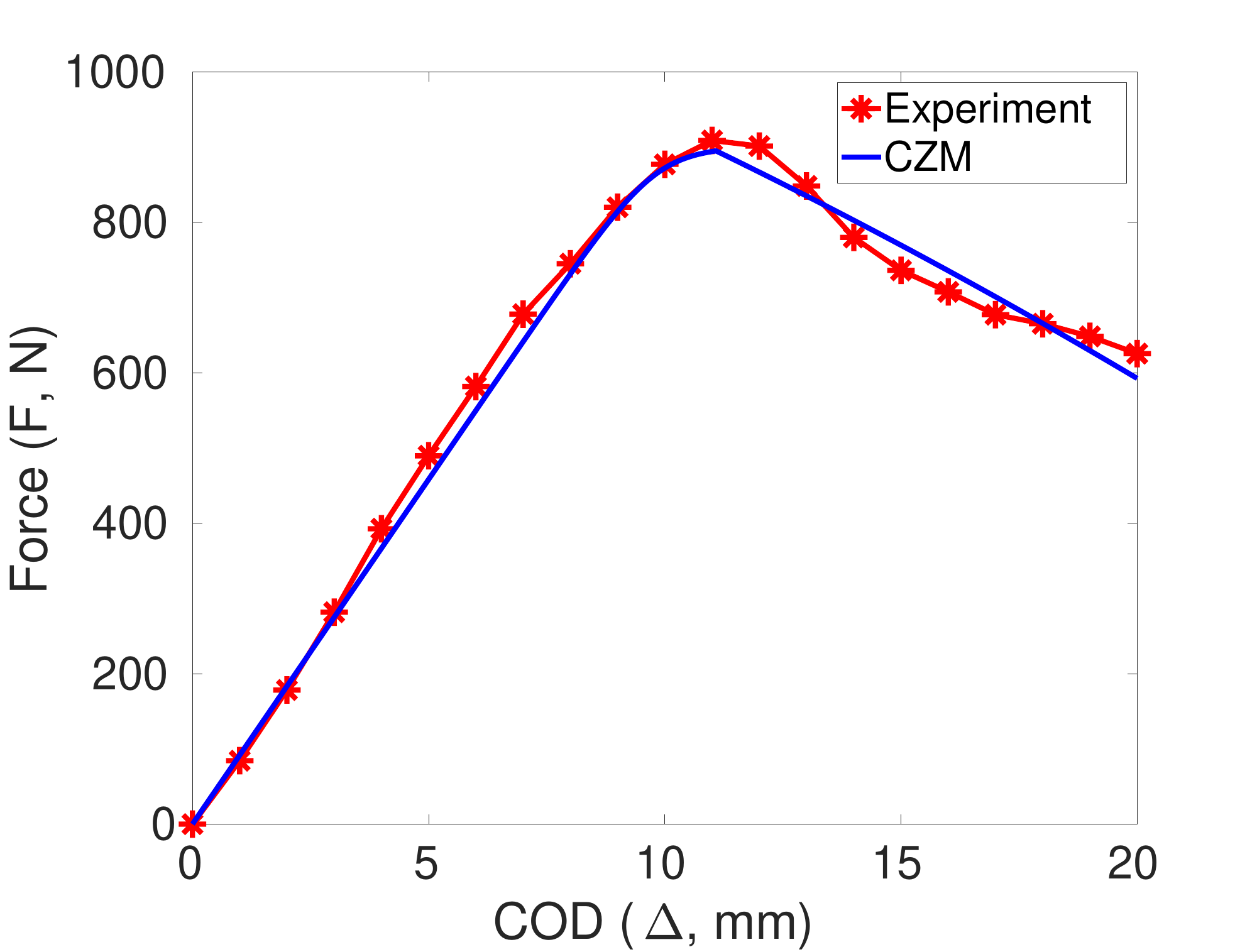}}
    \caption{Evaluation of computational model with posterior means  of the unknown parameters $\bm \theta$ and the experimental points used for calibration}
    \label{fig:bayes_comp}
\end{figure}

Figure \ref{fig:bayes_comp} shows the evaluation of the computational model after calibration. It is seen from the results that the parameters obtained from Bayesian calibration correctly predict the experimental response with reasonable overall error between the model evaluation and the experimental response. 

\begin{figure}[htpb]
    \centering
    \includegraphics[width=0.7\linewidth]{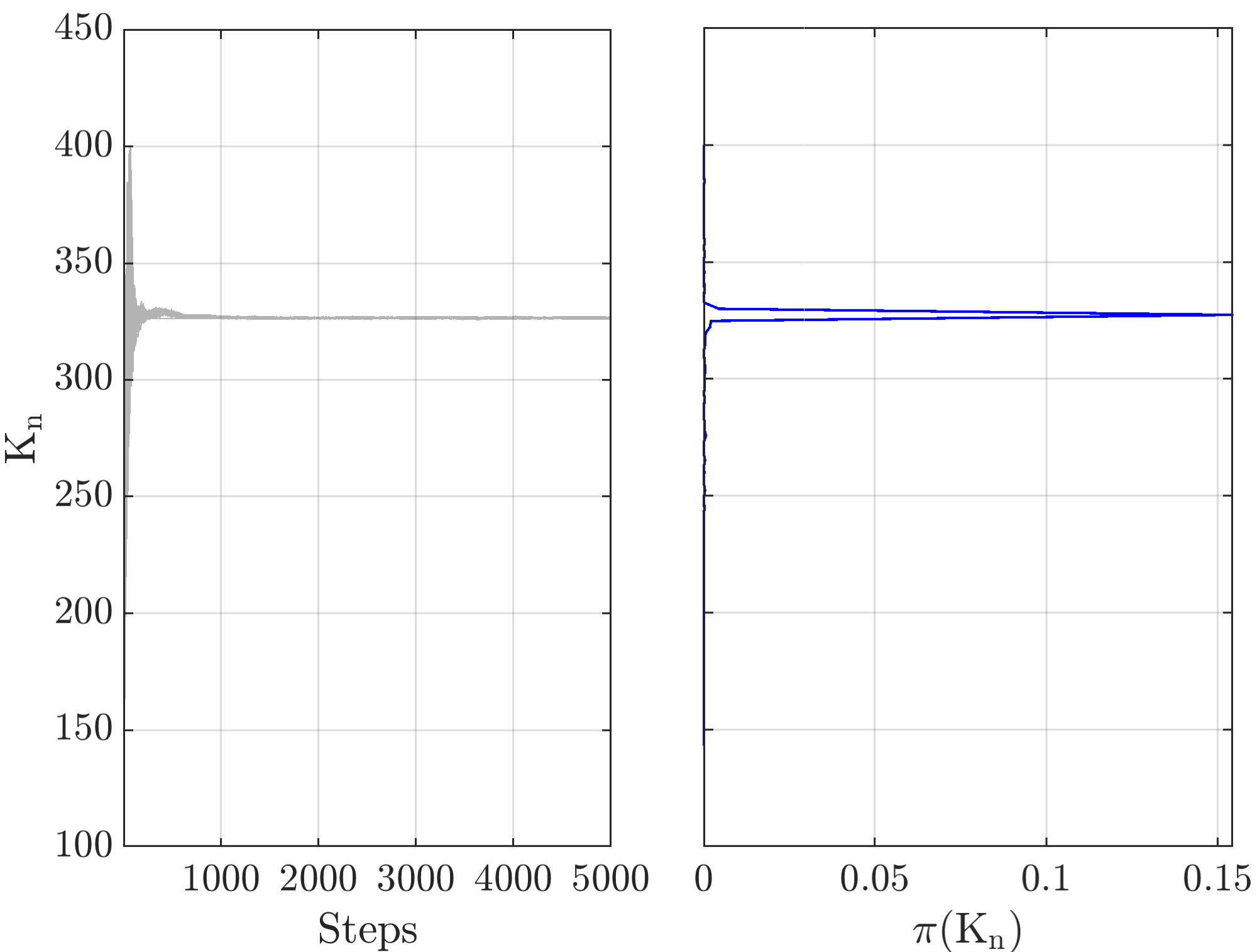}
    \caption{Trace and density plots for the $K_N$ parameter at 5 mm/min strain rate}
    \label{fig:trace_plot}
\end{figure}

Trace and density plots for the 100 randomly initialized MCMC chains are shown in Figure \ref{fig:trace_plot}. The density plots show multiple peaks for the $K_N$ parameter indicating the nonuniqueness of parameter values. This justifies the need for obtaining the posterior distribution of parameters instead of a single deterministic value.

Further analysis on the posterior distributions of the unknown parameters is given in \ref{sec:post_par}. 
Although the Bayesian calibration improves the CZM predictions, there is still a significant amount of discrepancy between the experimental response and the predictions as seen in Figure \ref{fig:bayes_comp}. To further improve the model's prediction we learn a discrepancy function as described in the following section. 

\section{Discrepancy function} \label{sec:disc}
\label{sec:disc_fn}
The discrepancy in predicting the experimental response is an important source of uncertainty in computational models such as the CZM. This discrepancy can arise due to missing physics, incorrect assumptions, numerical approximations, and/or other inaccuracies of the computational model. To account for this discrepancy, which is a source of uncertainty, a discrepancy function is introduced. This discrepancy function together with the computational model will provide better predictions of the experimental response. The functional form of the discrepancy function varies with varying applications. In this work, we learn the discrepancy function using a Gaussian process model $\mathbf{Y}_\delta(\mathbf{x}) \sim GP( \bm \mu_{\delta},  \bm \Sigma_{\delta})$. A brief overview of Gaussian process models is presented in the following.

\subsection{Gaussian Process (GP) models}
A GP is a collection of random variables where each random variable and any finite linear combination of these random variables are distributed normally. A GP model is a stochastic model for the prediction of output variable distributions which assumes that the output of the model $\mathbf{Y}(\mathbf{x})$ is a realization of a GP.
\begin{equation}
    \mathbf{Y}(\mathbf{x}) \sim GP(\bm{\beta}^T \bm{f}(\bm x), {\sigma}^2 \bm{R(x,x';\theta)})
\end{equation}
where, $\bm{\beta}^T \bm{f}(\bm x)$ is the mean of the GP, where $\bm{f}(\bm x) = \{f_i; i=1,...,P\}$  is a array of $P$ arbitrary functions and  $\bm{\beta}$ is the array of their coefficients. $\bm{R(x,x';\theta)}$ is the correlation function with hyperparameters $\bm \theta$ and $\sigma^2$ is a constant representing the variance.\\

With this assumption, the prediction $\hat{\bm Y}(\bm x)$ at a new input point $\bm x$, given a set of known model responses $\overline{\bm Y} =  \{\bm y^{(1)},...,\bm y^{(N)}\}$  at input points $\overline{\bm X} =  \{\bm x^{(1)},...,\bm x^{(N)}\}$, has a joint Gaussian distribution defined by \cite{rasmussen2006gaussian,santner2003design}
\begin{align}
    \begin{Bmatrix}
        \hat{\bm Y}(\bm x)\\
        \overline{\bm Y}
    \end{Bmatrix} \sim N_{N+1} 
    \begin{pmatrix}
        \begin{Bmatrix}
            \bm {f}^T (\bm x) \bm \beta \\
            \bm F \bm \beta
        \end{Bmatrix} + \sigma^2
        \begin{Bmatrix}
            1 & \bm {r}^T (\bm x) \\
            \bm {r}(\bm x) & \bm R
        \end{Bmatrix}
    \end{pmatrix}
\end{align}
where \\
    $\bm F =  F_{ij} = f_j(\bm x^{(i)});\;\; i=1,...,N; j=1,...,P$  \\
    $\bm r = r_i = R(\bm x,\bm x^{(i)}, \bm \theta);\;\; i = 1,...,N$\\
    $\bm R = R_{ij} = R( \bm x^{(i)}, \bm x^{(j)};\bm \theta);\;\; i,j = 1,...,N $\\
The mean and the variance of the prediction  $\hat{\bm Y}(\bm x)$ can be estimated as,
\begin{align}
    \mu_{\hat{\bm Y}}(\bm x) &= \bm f^T(\bm x) \hat{\bm \beta} + \bm r^T(\bm x) \bm R^{-1} \left(\overline{\bm Y} - \bm F \hat{\bm \beta}\right)\\
    \sigma_{\hat{\bm Y}}^2 (\bm x) &= \sigma^2 \left( 1 - \bm r^T(\bm x)\bm R^{-1} \bm r(\bm x) + \bm u^T (\bm x)(\bm F^T \bm R^{-1} \bm F)^{-1} \bm u(\bm x)\right)
\end{align}
where,
\begin{align*}
 &\hat{\bm \beta} = (\bm F^T \bm R^{-1} \bm F)^{-1} \bm F^T \bm R^{-1} \overline{\bm Y}; &\bm u(\bm x) =  \bm F^T \bm R^{-1} \bm r (\bm x) - \bm f (\bm x)   
\end{align*}
In this work, we use this Gaussian process regression model to learn the discrepancy function.
\subsection{Results of the discrepancy function}
  The mean of the posterior distributions obtained from Bayesian calibration (Eqn. \ref{eq:bayes}) is used to evaluate the computational model for the three strain rates. The discrepancy ($\delta$) is calculated as the difference between the experimental response and the output of the computational model at a given input ($\bm X$). Twenty uniformly spaced points are selected on the load-displacement curve to learn the discrepancy function. Further details on the convergence of error with the number of points are given in \ref{app:conv}. Three Gaussian processes, one for each strain rate, are learned for the input and the discrepancy. A zeroth-order polynomial is taken as the mean of the Gaussian process and an ellipsoidal correlation function is used. The hyperparameters of the Gaussian process are obtained by minimizing the cross-validation error. A hybrid Genetic algorithm method is used as the optimization method to obtain the hyperparameters. In making any new predictions, the output of the computational model is corrected using this discrepancy function. 

\begin{figure}[ht]
    \centering
    \includegraphics[width = 0.8\linewidth]{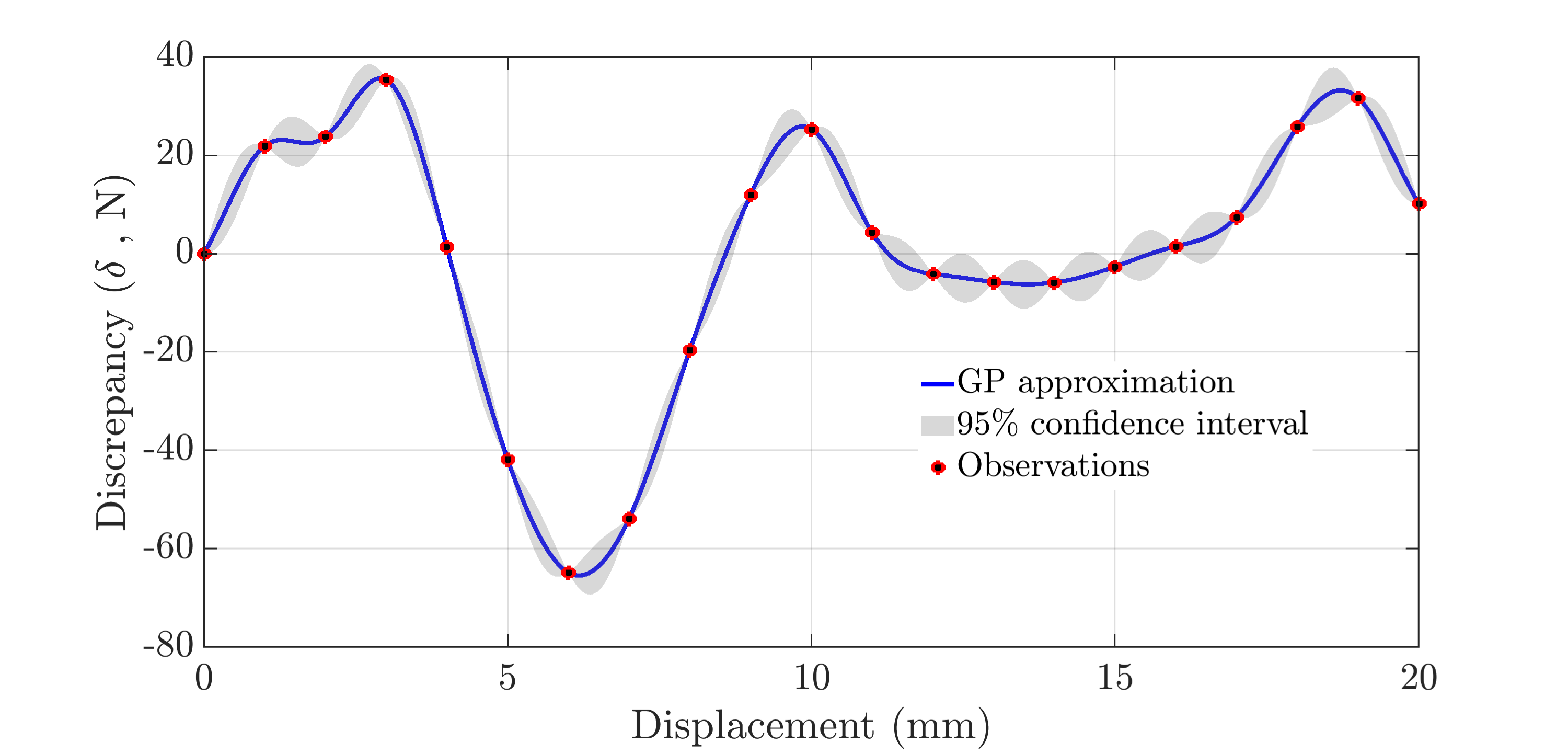}
    \caption{Gaussian process for the discrepancy in the prediction of 5 mm/min strain rate.}
    \label{fig:disc_gp}
\end{figure}

The Gaussian process learned for the discrepancy between the experimental response and the model prediction is shown in Figure \ref{fig:disc_gp}. It is seen that the GP predicts the discrepancy accurately and with minimum uncertainty at the training data points. The nature of the discrepancy across the input also suggests that a simple polynomial model cannot learn this discrepancy and a GP is essential in this case.

\begin{figure}[ht]
    \centering
    \subfigure[5 mm/min strain rate]{\label{fig:disc_a}\includegraphics[width = 0.45\linewidth]{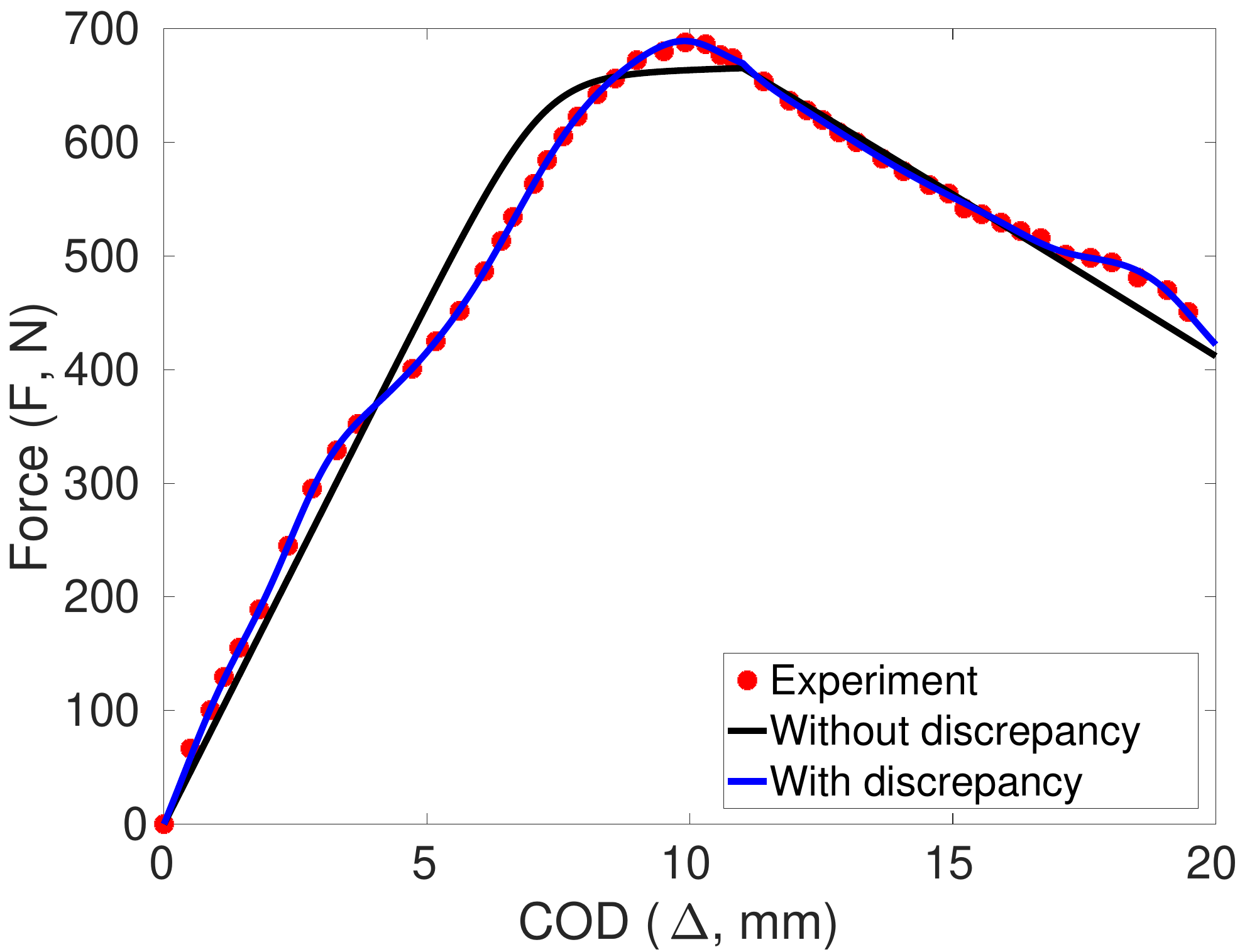}}
    \hfill
    \subfigure[50 mm/min strain rate]{\label{fig:disc_b}\includegraphics[width = 0.45\linewidth]{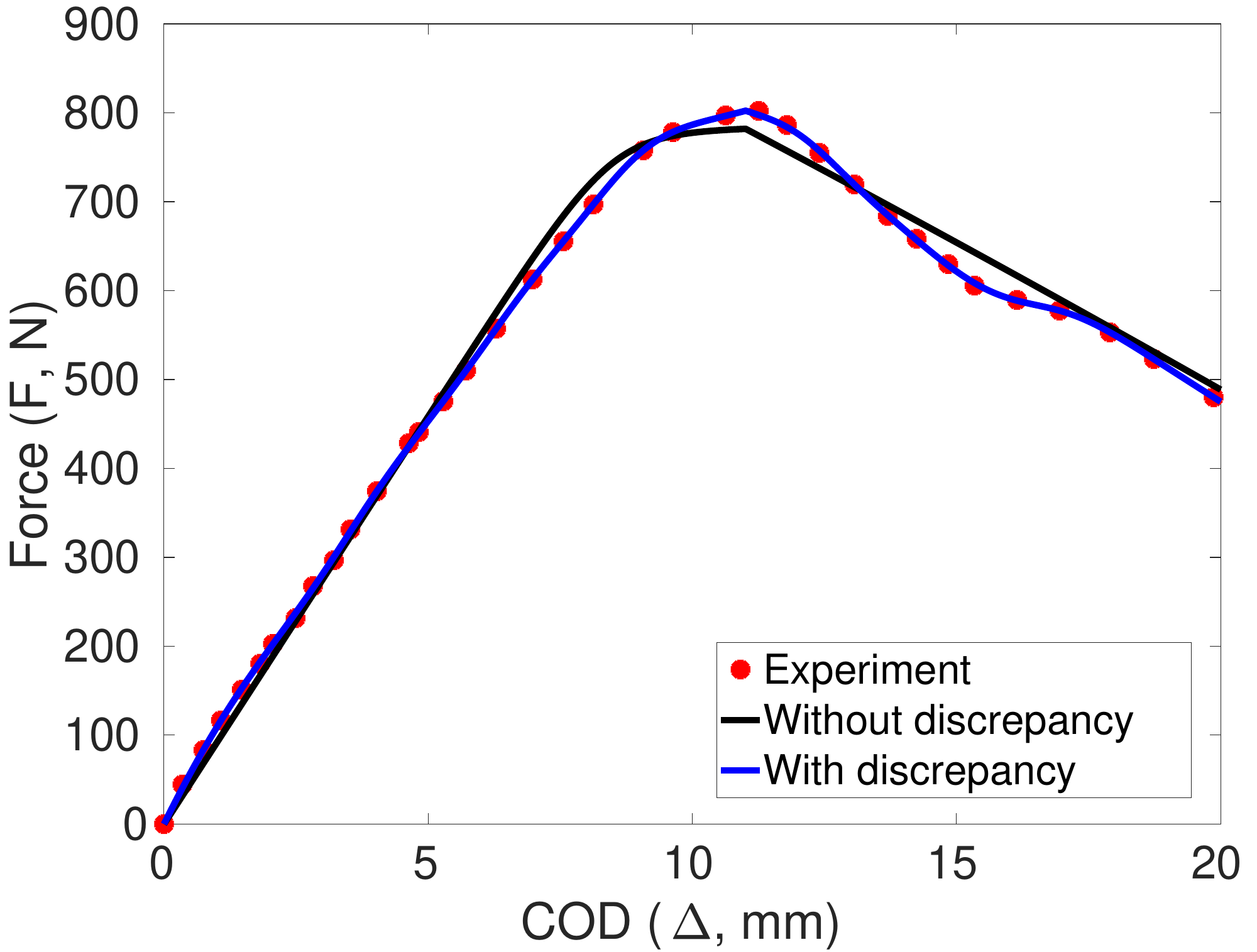}}
    \subfigure[500 mm/min strain rate]{\label{fig:disc_c}\includegraphics[width = 0.45\linewidth]{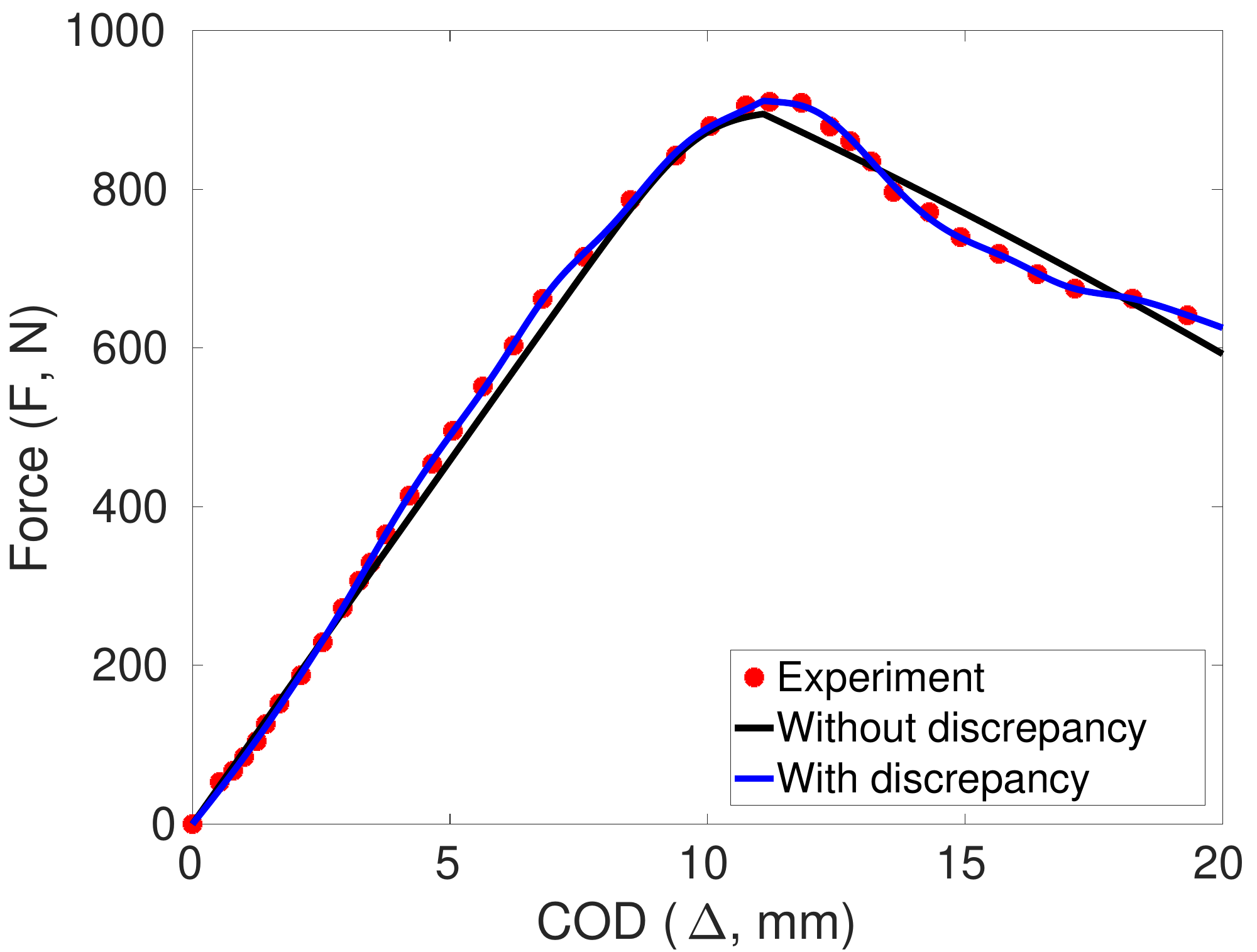}}
    \caption{Model evaluation with and without discrepancy function}
    \label{fig:disc}
\end{figure}

 The results of the computational model with the discrepancy function are presented in Figure \ref{fig:disc}. The results show that after the introduction of a discrepancy function, the model's  prediction of the experimental response improved significantly for all three strain rates.
To quantify the improvement in the CZM's predictions with the addition of the discrepancy function,  the percentage error is evaluated as $err = \frac{\sqrt{\sum_i (y^e_i - y^p_i)^2}}{\sqrt{\sum_i (y^e_i)^2}} \times 100$ and provided in table \ref{tab:per_error} below \footnote{$y^p$ is the predictions of the CZM with or without the discrepancy function. These errors are calculated for data points which are not used to train the GP for discrepancy function.}. 

\begin{table}[htpb]
    \centering
    \begin{tabular}{ccc} \toprule
         Strain rate& \% Error CZM  & \% Error CZM with $\bm \delta$ \\ \midrule
         5 mm/min & 22.60& 6.52  \\ 
         50 mm/min & 17.28 & 5.05 \\
         500 mm/min & 17.98 & 6.86 \\ \bottomrule
    \end{tabular}
    \caption{Percentage error in the CZM predictions with and without the discrepancy function}
    \label{tab:per_error}
\end{table}

\section{Uncertainty Quantification} \label{sec:uq}
A computational model, such as the CZM, is a mathematical representation of a physical phenomenon. Physical phenomenons have natural variability associated with them which are referred to as aleatoric uncertainties. Further, modeling of these physical phenomena introduces additional uncertainties which may be a result of limited measurement data, imprecise measurement, solution approximations, unknown model parameters, and model assumptions. These uncertainties are referred to as epistemic uncertainties. The aleatoric and epistemic uncertainties need to be modeled and quantified to better understand and confidently use the computational models to obtain predictions.

\begin{figure}[htpb]
    \centering
    \includegraphics[width=0.7\linewidth]{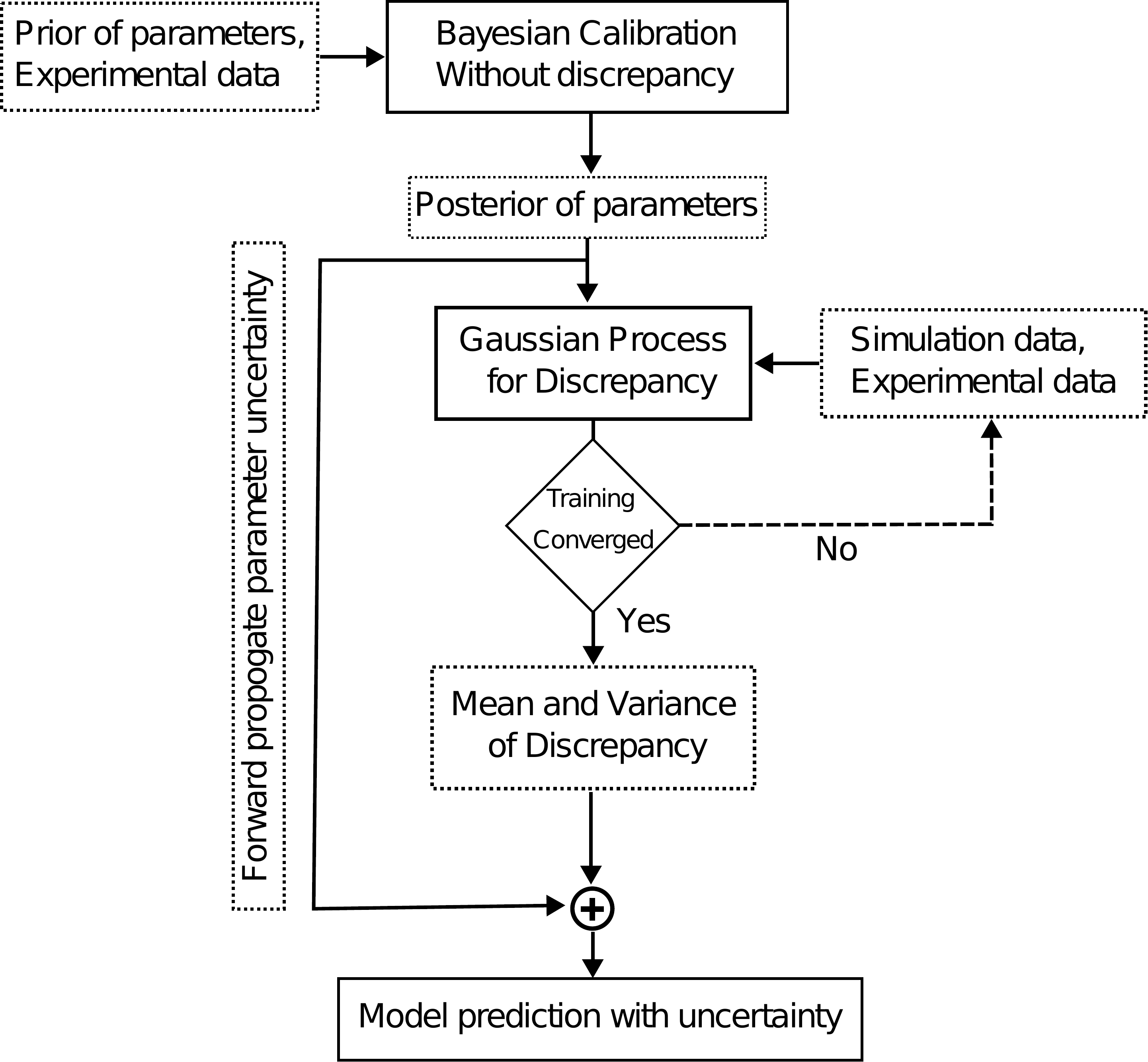}
    \caption{Uncertainty quantification framework}
    \label{fig:flow_uq}
\end{figure}

Bayesian calibration quantifies the parameter uncertainty and measurement errors. These uncertainties can be propagated to the computational model through the posterior distribution of the calibrated parameters. The discrepancy function quantifies the uncertainties due to the model assumptions or missing physics. Hence, the total uncertainty in the prediction can be quantified as the sum of uncertainty in the computational model and the uncertainty in the discrepancy function. The steps involved in quantifying uncertainties in the predictions of the CZM are summarised in the flow chart in Fig.~\ref{fig:flow_uq}.

To quantify the overall uncertainty,  the uncertainties in the model parameters are propagated through the CZM model by sampling the posterior distribution of parameters and evaluating the model at each of these samples. The variance in the prediction of these samples ($\bm \Sigma_{c} (\bm X, \bm \theta)$) is an estimate of the forwarded propagated parameter uncertainty. Therefore, the total uncertainty is  
\begin{equation}
    \bm \Sigma_{pred}(\bm X) = \bm \Sigma_{c} (\bm X, \bm \theta) + \bm \Sigma_{\delta} (\bm X)
    \label{eq:UQ}
\end{equation}

In a predictive setting, any new prediction $\bm Y_{pred}$ for the CZM can be evaluated as,
\begin{equation}
    \bm Y_{pred} = Y_c(\bm X, \bm \theta^*) + \bm \mu_{\bm \delta}
\end{equation}
Where $Y_c(\bm X, \bm \theta^*)$ and $ \bm \mu_{\bm \delta}$ are the model's prediction at the calibrated parameters and the mean of the GP for discrepancy function respectively. The diagonal values of $\bm \Sigma_{pred}$, denoted as $\bm \sigma_{pred}^2$ are used in determining the confidence intervals of predictions as,  
\begin{equation}
    \bm Y_{pred} \in \left[\bm \mu_{pred} - \Phi^{-1}\left(1-\frac{\alpha}{2}\bm \sigma_{pred}^2\right) , \bm \mu_{pred} + \Phi^{-1}\left(1-\frac{\alpha}{2}\bm \sigma_{pred}^2 \right)\right]
\end{equation}
with probability $1-\alpha$. Where $\Phi(.)$ is the cumulative distribution function of the Gaussian distribution. 

\subsection{Results of uncertainty quantification}
The posterior distribution of the unknown model parameters, $p(\bm \theta | \bm d)$, are sampled to propagate the parameter uncertainty through the computational model. 1000 samples of size $1000 \times 8$  are generated from the posterior distribution of parameters learned using Bayesian calibration.  The computational model is evaluated using these samples and the variance in the output provides us with the uncertainty of the computational model $\bm \Sigma_{Y^c} (\bm X, \bm \theta)$.  With the uncertainty from the discrepancy function, $\bm \Sigma_{\delta} (\bm X)$, the total uncertainty of the prediction can be evaluated from Eqn. \eqref{eq:UQ}

\begin{figure}[htpb]
    \centering
    \subfigure[5 mm/min strain rate]{\label{fig:uq_a}\includegraphics[width = 0.49\linewidth]{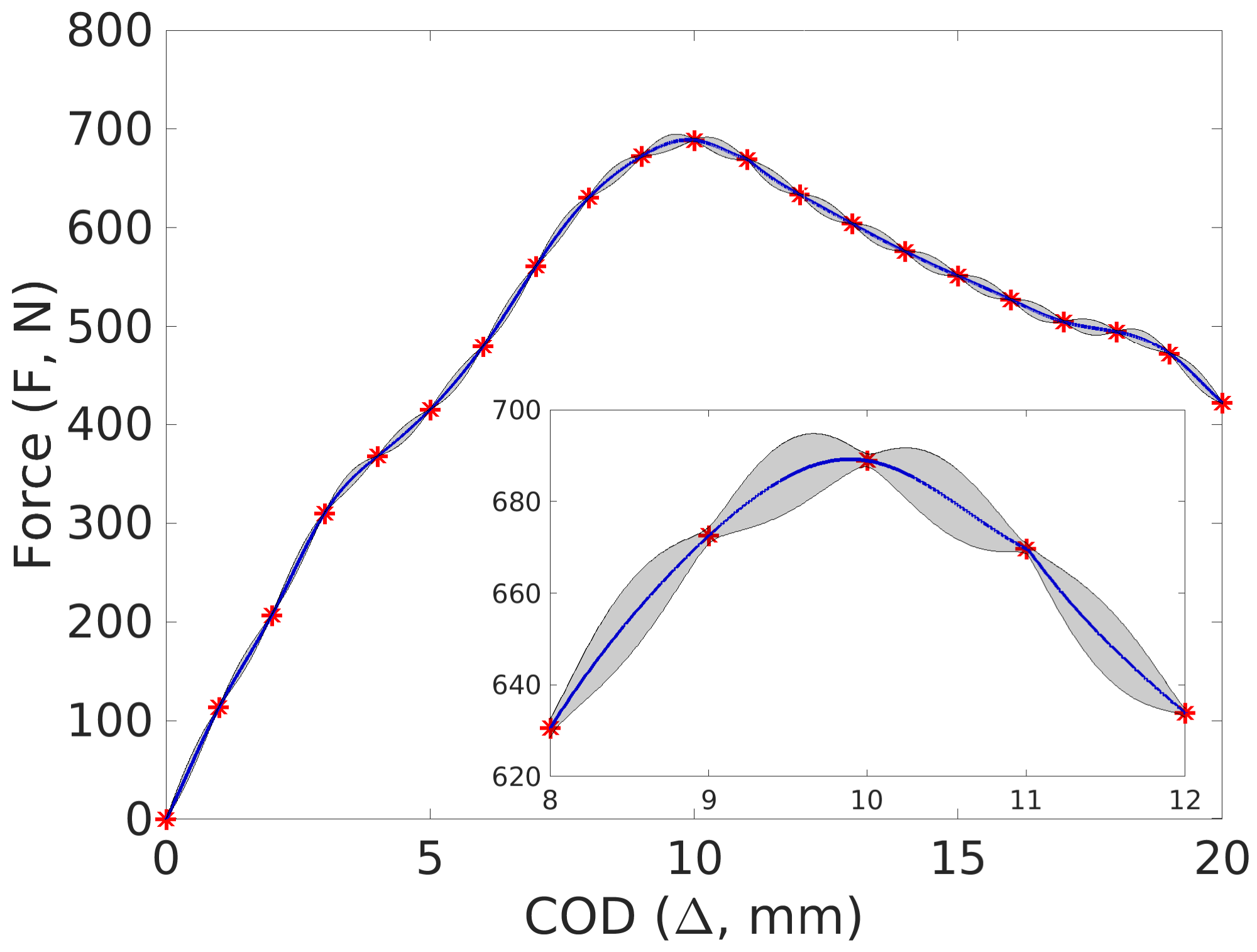}}
    \hfill
    \subfigure[50 mm/min strain rate]{\label{fig:uq_b}\includegraphics[width = 0.49\linewidth]{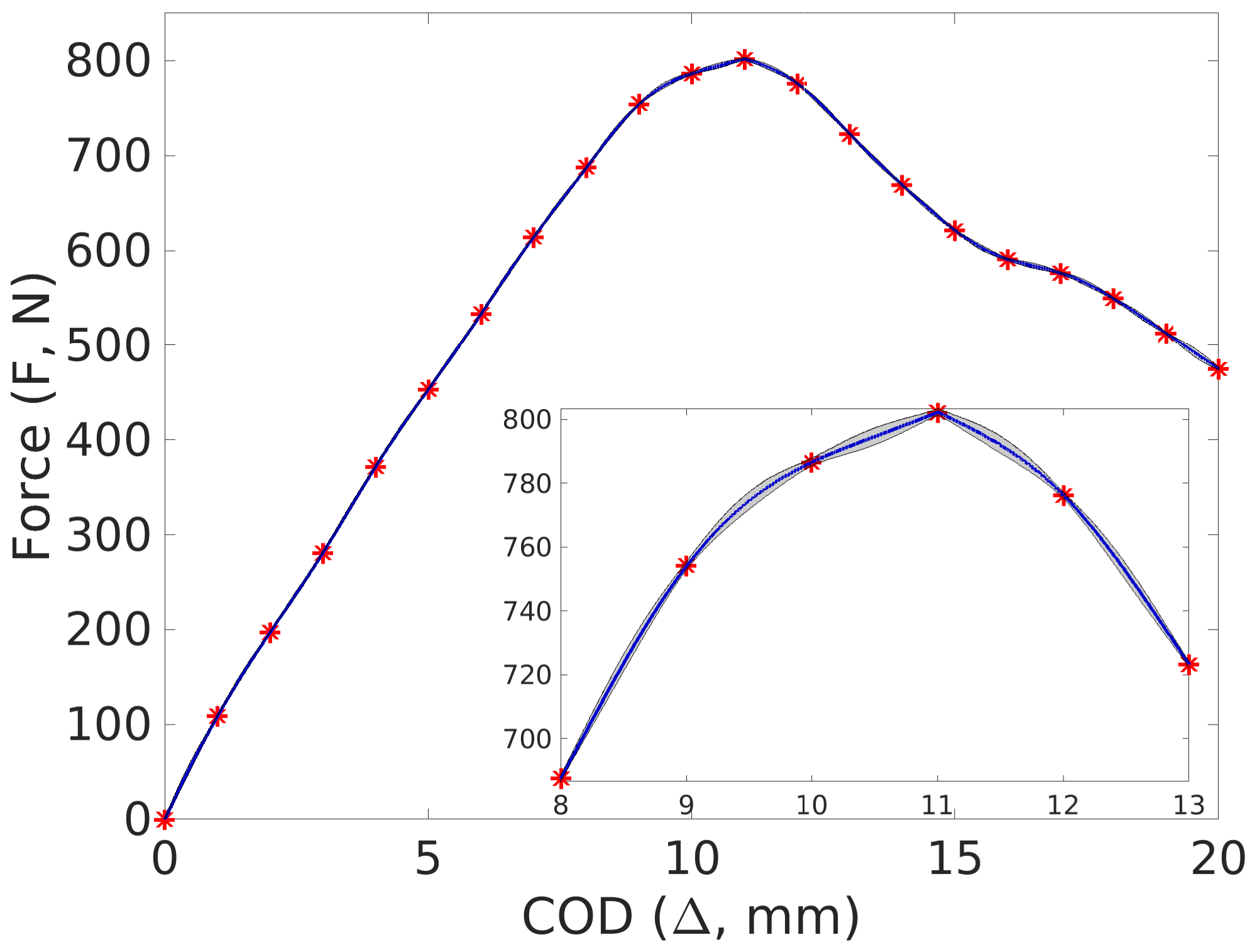}}
    \subfigure[500 mm/min strain rate]{\label{fig:uq_c}\includegraphics[width = 0.5\linewidth]{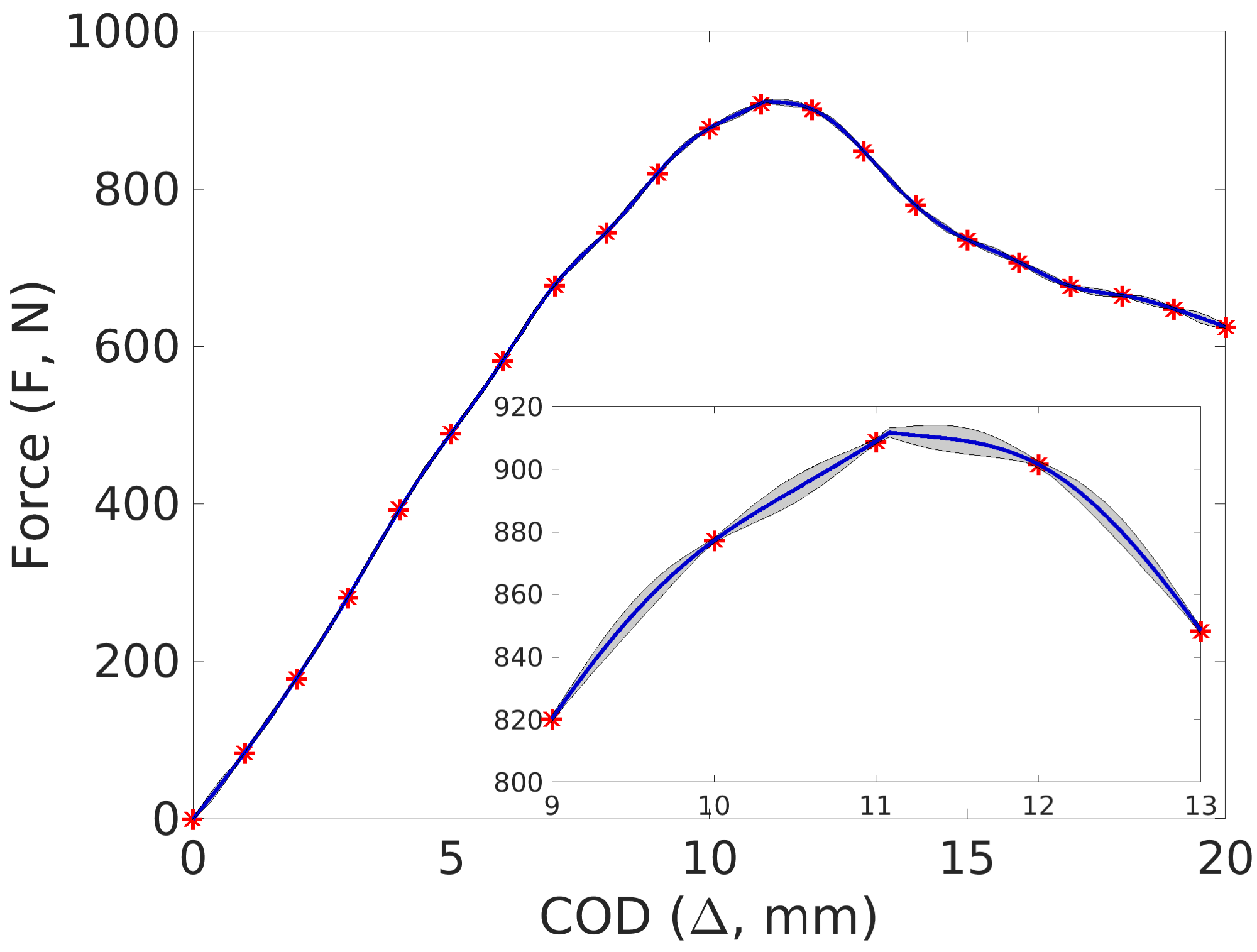}}
    \caption{Model predictions with $\pm 3\sigma$ confidence intervals}
    \label{fig:uq}
\end{figure}

Uncertainties are presented as confidence intervals in figure \ref{fig:uq}. It is seen that with the inclusion of the discrepancy function, all the experimental data lies well within the 95\% confidence interval.

\section{Sensitivity Analysis} \label{sec:sensitivity}
A sensitivity analysis provides a better understanding of the input-output relationship in the computational model. The contribution of individual input parameters to the overall uncertainty of the output of the computational model can be studied from a sensitivity analysis. This can also help in simplification of a stochastic model by assuming the less sensitive random parameters to be deterministic. 

A number of methods have been developed to perform sensitivity analysis in the literature \cite{iooss2015review}. These methods can be broadly classified as 1) local methods: which involve the study of small input perturbations around nominal values on the model output 2) global methods: which consider the range of the whole input domain. One such global method for sensitivity analysis is the Sobol' indices or the analysis of variance \cite{sobol1993sensitivity}. 

In this method, the total variance of a model is decomposed into the variance of the summands as,
\begin{equation*}
    \text{Var}(Y) = \sum_{i=1}^d V_i + \sum_{i<j}^d V_{ij} + ... + V_{12...d}
\end{equation*}
where,
\begin{align*}
    &V_i = \text{Var}_{X_i}[\mathbb{E}_{X_{\sim i}}(Y|X_i)]; & V_{ij} = \text{Var}_{X_{ij}}[\mathbb{E}_{X_{\sim ij}}(Y|X_{ij})] - V_i - V_j
\end{align*}
and so on. The notation $X_{\sim i}$ indicates the set of all variables except $X_i$ and $\mathbb{E}$ is the expectation.\\
The first-order indices are given by,
\begin{equation*}
    S_i = \frac{V_i}{\text{Var}(Y)}
\end{equation*}
and the total-order indices are given by,
\begin{equation*}
    S_{Ti} = \frac{\mathbb{E}_{X_{\sim i}}[\text{Var}_{X_i}(Y|X_i)]}{\text{Var}(Y)}
\end{equation*}

\begin{equation*}
    f(x_1,...,x_M) = f_0 + \sum_{i=1}^M f_i(x_i) + \sum_{1 \le i < j \le M} f_{ij}(x_i,x_j) + ... + f_{1,2,...,M} (x_1,...,x_M)
\end{equation*}
where the following conditions hold:
\begin{enumerate}
    \item $f_0$ is equal to the expected value of $f(\boldsymbol{X})$.
    \item Integrals with respect to their own variables is zero.  $$\int_0^1 f_{i_1,...,i_s}(x_{i_1},...,x_{i_s}) dx_{i_k}  = 0 \text{ for } 1 \le k \le s.$$
\end{enumerate}
The summands are calculated as follows:
\begin{align*}
    f_0 &= \int_{D_X} f(x) dx,\\
    f_i(x_i) &= \int_0^1 ... \int_0^1 f(x) dx_{\sim i} - f_0,\\
    f_{ij}(x_i,x_j) &= \int_0^1 ... \int_0^1 f(x)dx_{\sim ij} - f_0 - f_i(x_i) - f_j(x_j),\\
\end{align*}

The total variance of $f(\boldsymbol{X})$ are computed as,
\begin{align*}
    D = \int_{D_X} f^2(x) dx - f^2_0
\end{align*}
and the partial variance are given by:
\begin{align*}
    D_{{i_1},...,{i_s}} = \int_0^1 ... \int_0^1 f^2_{{i_1},...,{i_s}}(x_{i_1},...,x_{i_s})dx_{i_1}...dx_{i_s} \text{  } 1 \le i_1 < ... < i_s \le M ; s = 1,...,M
\end{align*}

Now, the first and higher order sensitivity indices can be defined as,
\begin{equation*}
    S_{i_1,...,i_s} = \frac{D_{i_1,...,i_s}}{D}
\end{equation*}
which represents the contribution of each group of variables ${X_{i_1},...,X_{i_s}}$ to the total variance. The index with respect to one input variable $X_i$ is called the first-order Sobol's index. Multiple term indices $S_{ij}, i=j$ are called the higher order Sobol' indices. \\
The total index of input variable $X_i$ is the sum of all the Sobol' indices involving this variable: 
\begin{equation*}
    S_i^T = \sum_{\{i_1,...,i_s\}\supset i} S_{i_1,...,i_s}
\end{equation*}
\subsection{Results of the sensitivity analysis}
A sensitivity analysis for the unknown parameters is performed based on the peak load as the output of the forward model. The parameters are sampled from the prior distributions to perform the sensitivity analysis. Displacement Value at damage onset ($\delta^o$) is the most sensitive parameter for this forward model as seen in Fig.~\ref{fig:sensitivity}. Reference plastic strain $\gamma_o$ is the least sensitive in determining the peak load.
\begin{figure}[htpb]
    \centering
    \subfigure[First order Sobol' indices]{\label{fig:a}\includegraphics[width = 0.48\linewidth]{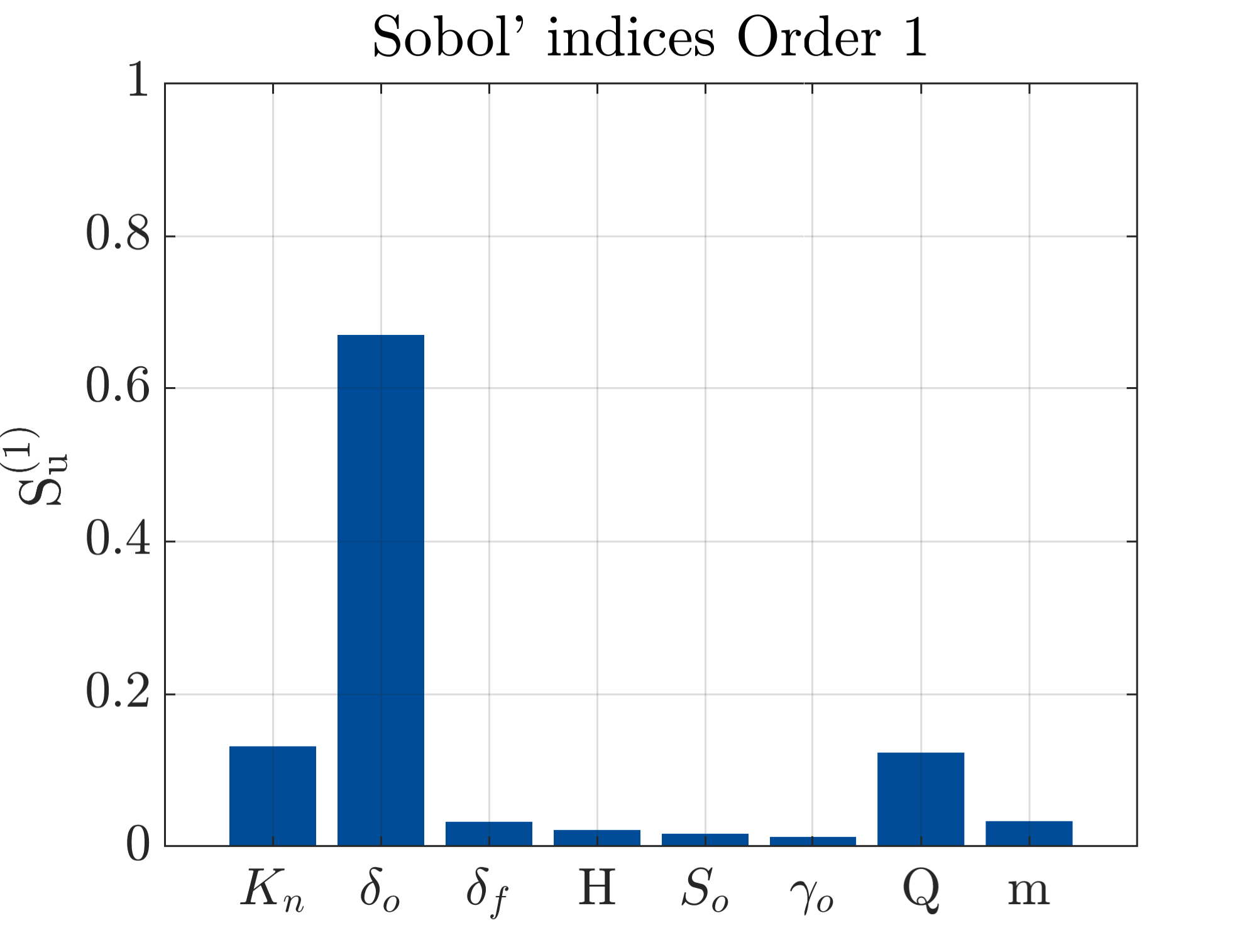}}
    \hfill
    \subfigure[Total Sobol' indices]{\label{fig:b}\includegraphics[width = 0.48\linewidth]{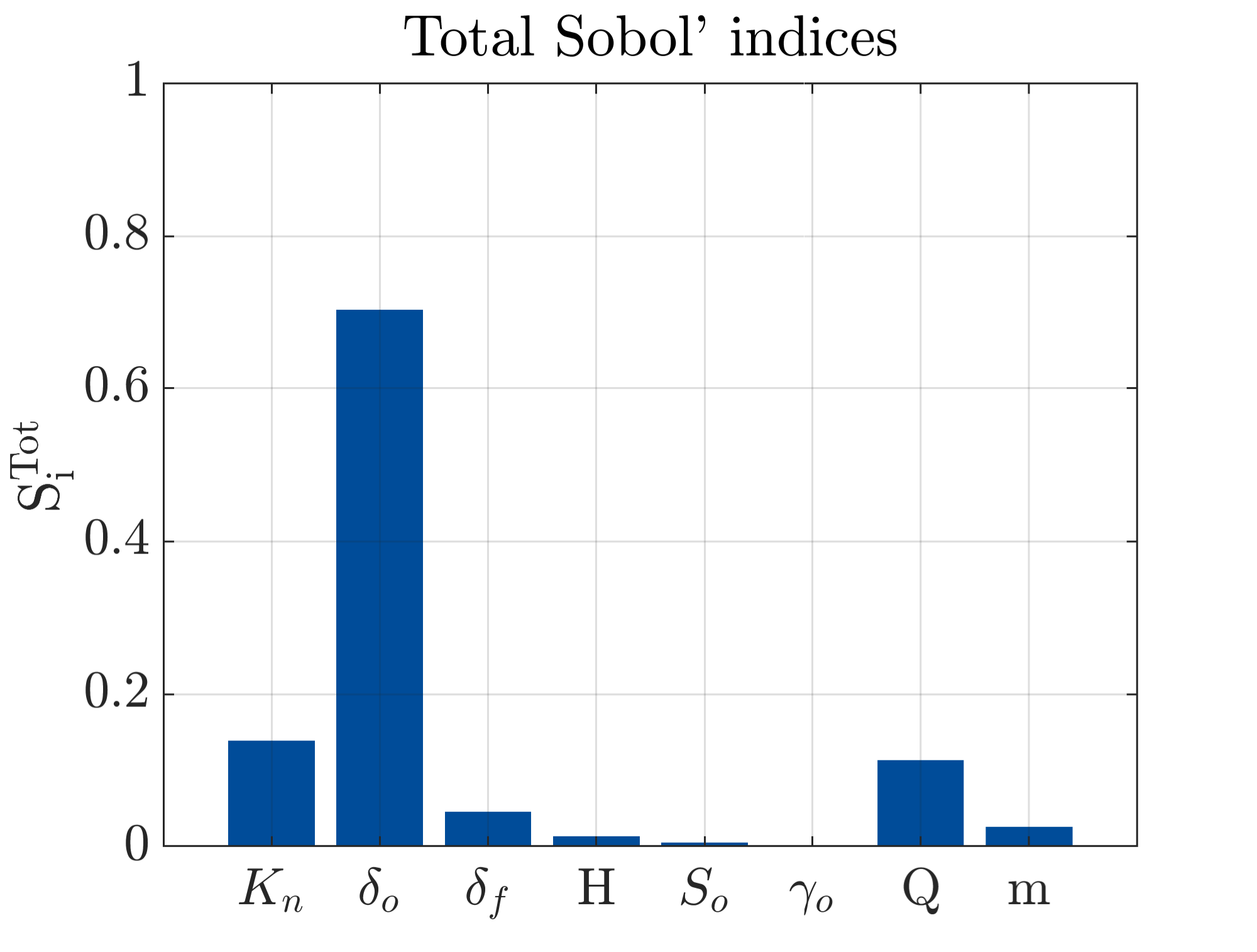}}
    \caption{Sobol indices for the calibration parameters}
    \label{fig:sensitivity}
\end{figure}


\section{Conclusions}
\label{sec:conc}
Calibration of the unknown parameters of CZM is carried out using a Bayesian approach. The Bayesian framework results in probability distributions for unknown parameters rather than a deterministic value. The variance of the probability distributions provides confidence in the calibrated values. This can be used as a tool to judge if more experimental data is necessary to improve the calibrated values. The discrepancy function is sequentially calibrated following the unknown parameters by fitting a Gaussian process. This function accounts for the difference between the predictions and experimental observations and helps in bridging the gap between the two, thereby providing predictions close to the observed data. Overall quantification of uncertainties is performed and the predictions are provided along with confident intervals. A sensitivity analysis to understand the input-output relationship is also carried out and the results are presented. 
\section*{Acknowledgments}
SG and PT acknowledge financial support from NSF (CMMI MoMS) grant number 1937983 and the U.S. Department of Energy, Office of Science, grant DE-SC0023432 and Finishing fellowship from the Michigan Tech graduate school. 
SG and PT acknowledge the supercomputing resources from the SUPERIOR computing facility at MTU and the Extreme Science and Engineering Discovery Environment (XSEDE), which is supported by the NSF grant number ACI-1548562. (Request number: MSS190003, MSS200004). TS acknowledges Mohammed R. Imam and Rishab Awasthi for their initial numerical work on the fracture model.


\bibliographystyle{elsarticle-num}
\bibliography{interface_Ref}

\appendix
\section{Analysis of the posterior of parameters} \label{sec:post_par}
In MCMC methods, trace plots serve as an important tool to diagnose the convergence of MCMC chains. Trace plots track the individual Markov chains during the optimization process. Trace plots for the 100 random initialized chains are presented in Fig.~\ref{fig:trace_1}. The plots for the parameters $K_N, \delta_o, \delta_f, H$ and $\gamma_o$ show good convergence. Given that the calibration is performed in high (eight) dimensions the plots for the parameters $S_o, Q$, and $m$ are reasonable. 
\begin{figure}[h!] \ContinuedFloat
    \centering
    \subfigure[]{\includegraphics[width=0.49\linewidth]{New_res/trace_kn.pdf}}
    \subfigure[]{\includegraphics[width=0.49\linewidth]{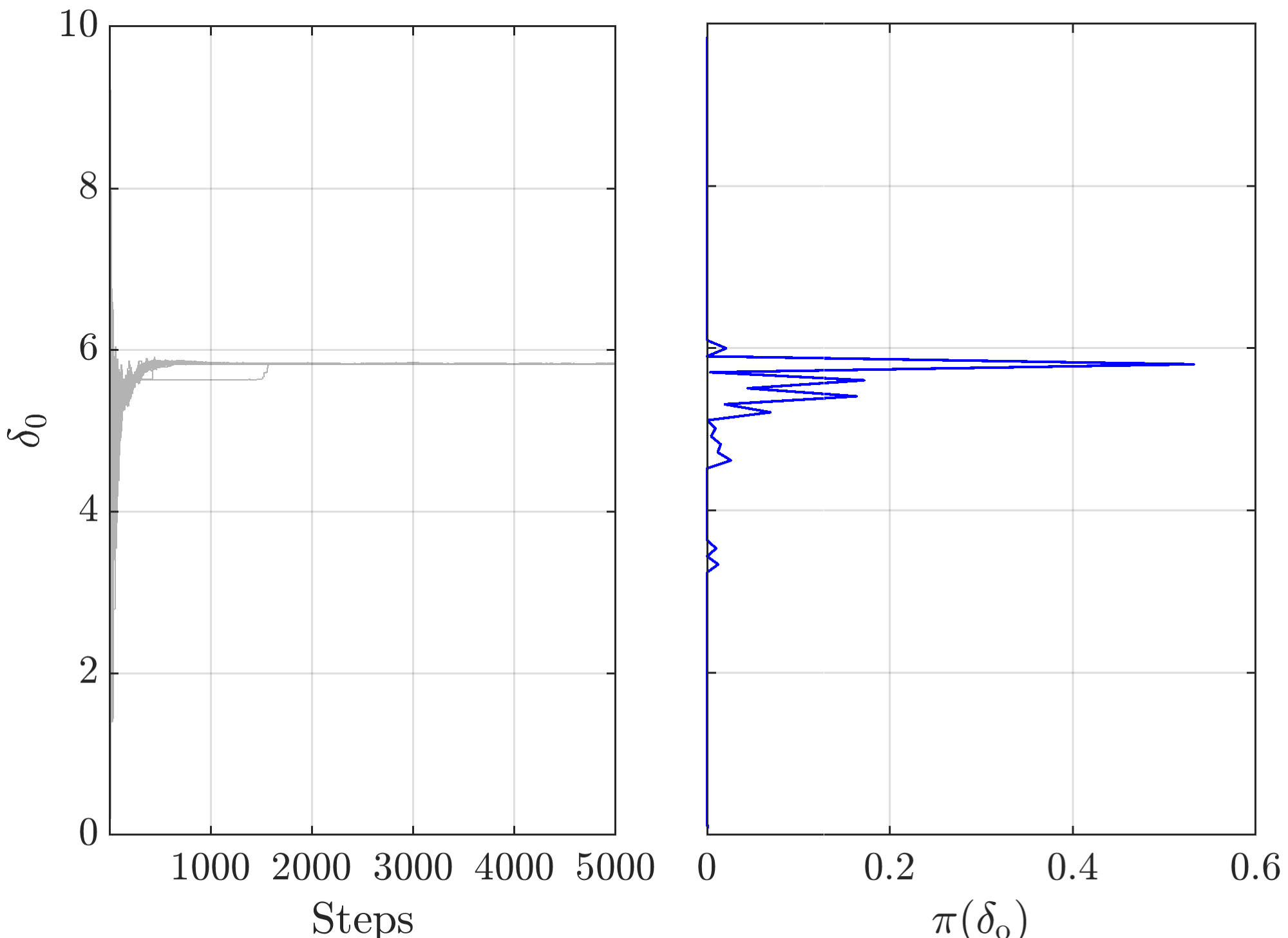}}
    \subfigure[]{\includegraphics[width=0.49\linewidth]{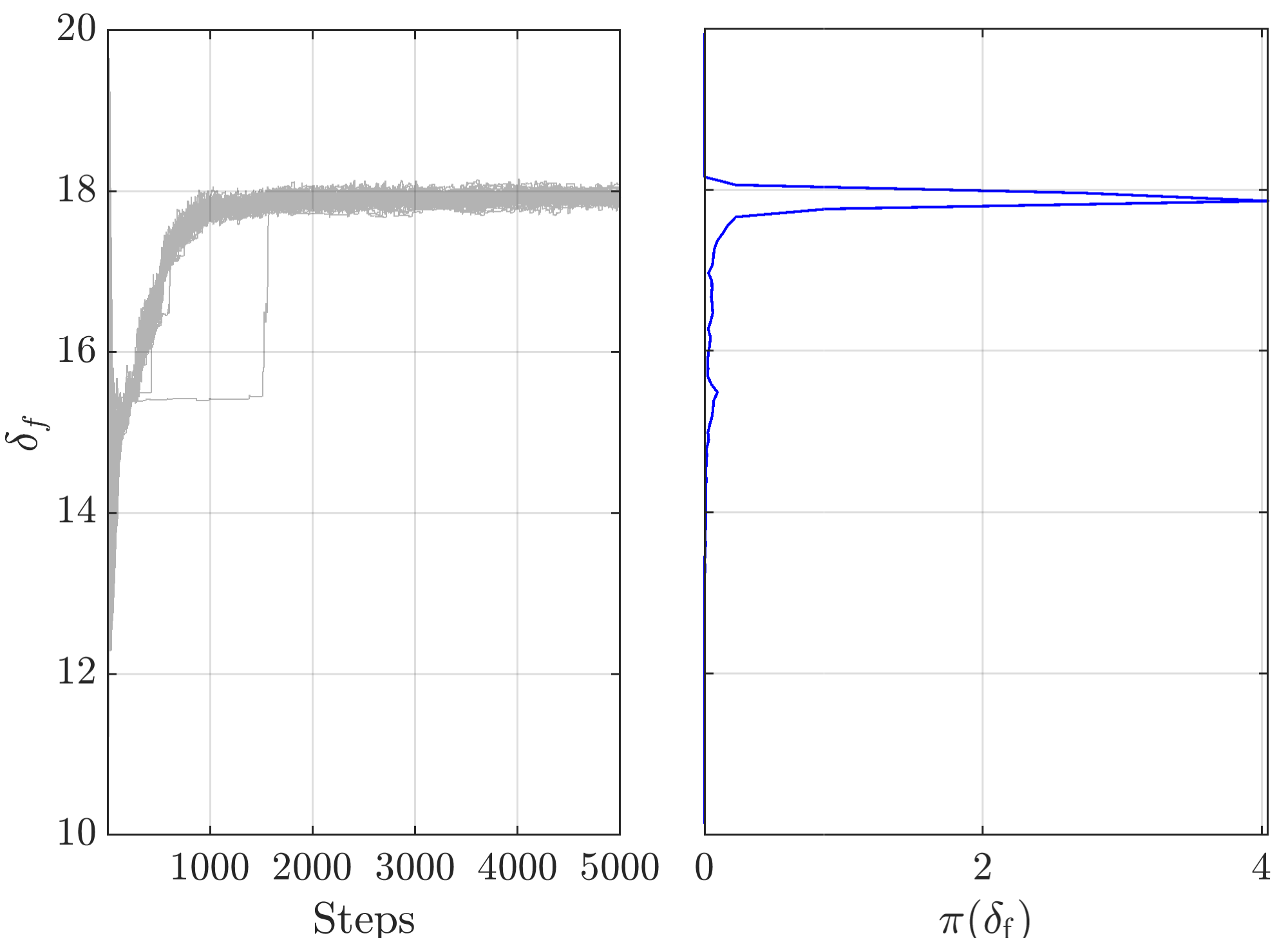}}
    \subfigure[]{\includegraphics[width=0.49\linewidth]{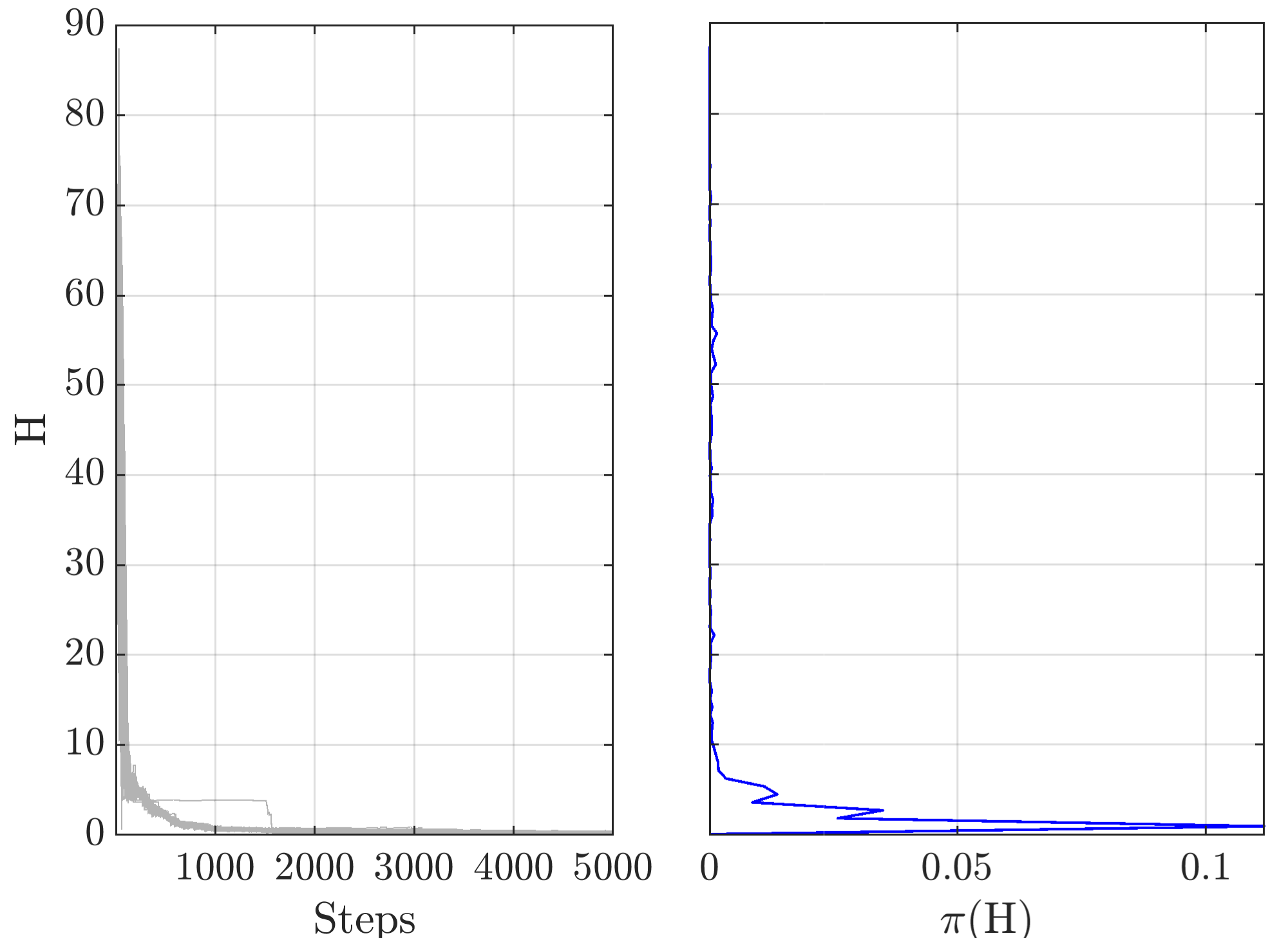}}
    \subfigure[]{\includegraphics[width=0.49\linewidth]{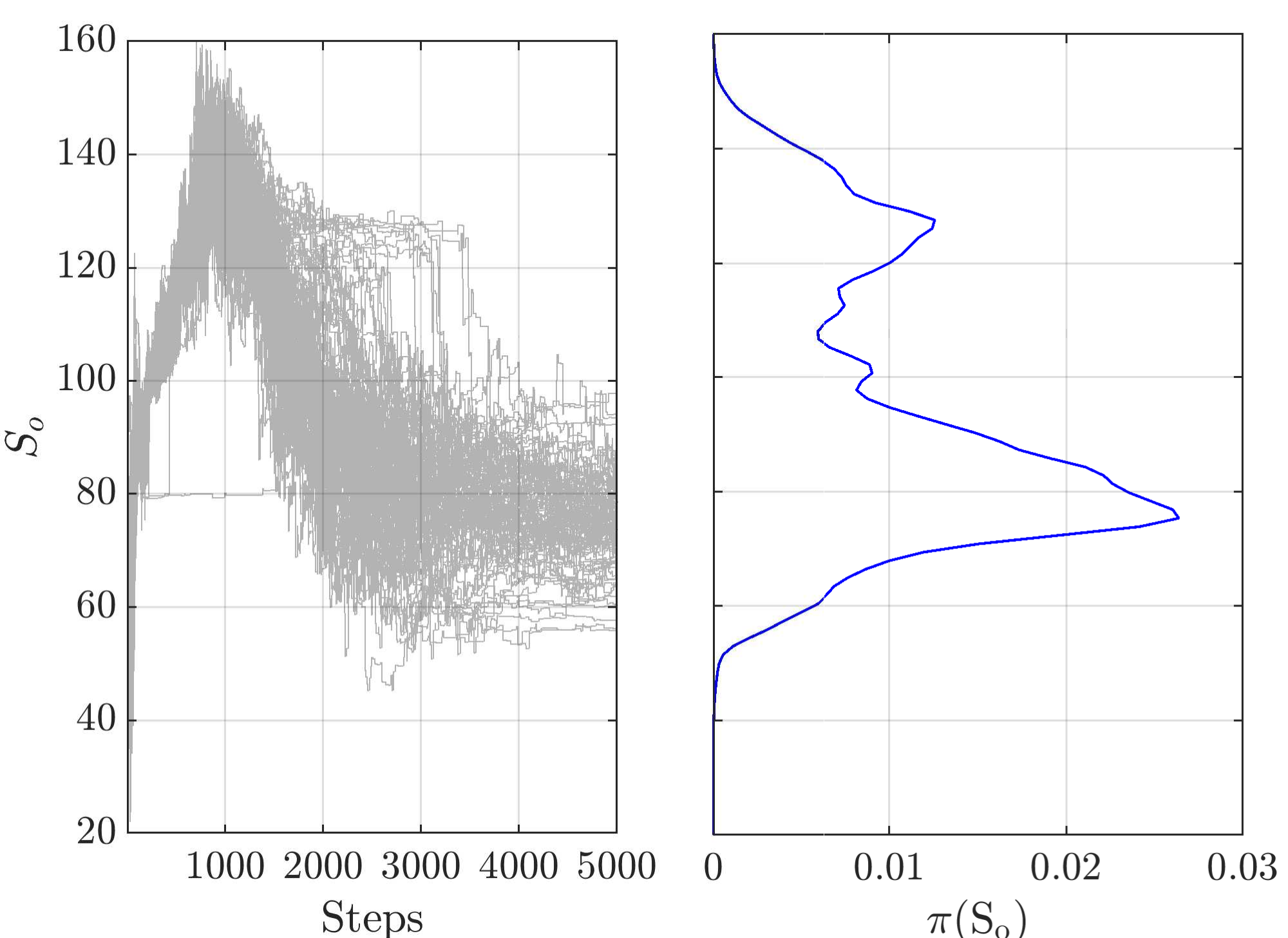}}
    \subfigure[]{\includegraphics[width=0.49\linewidth]{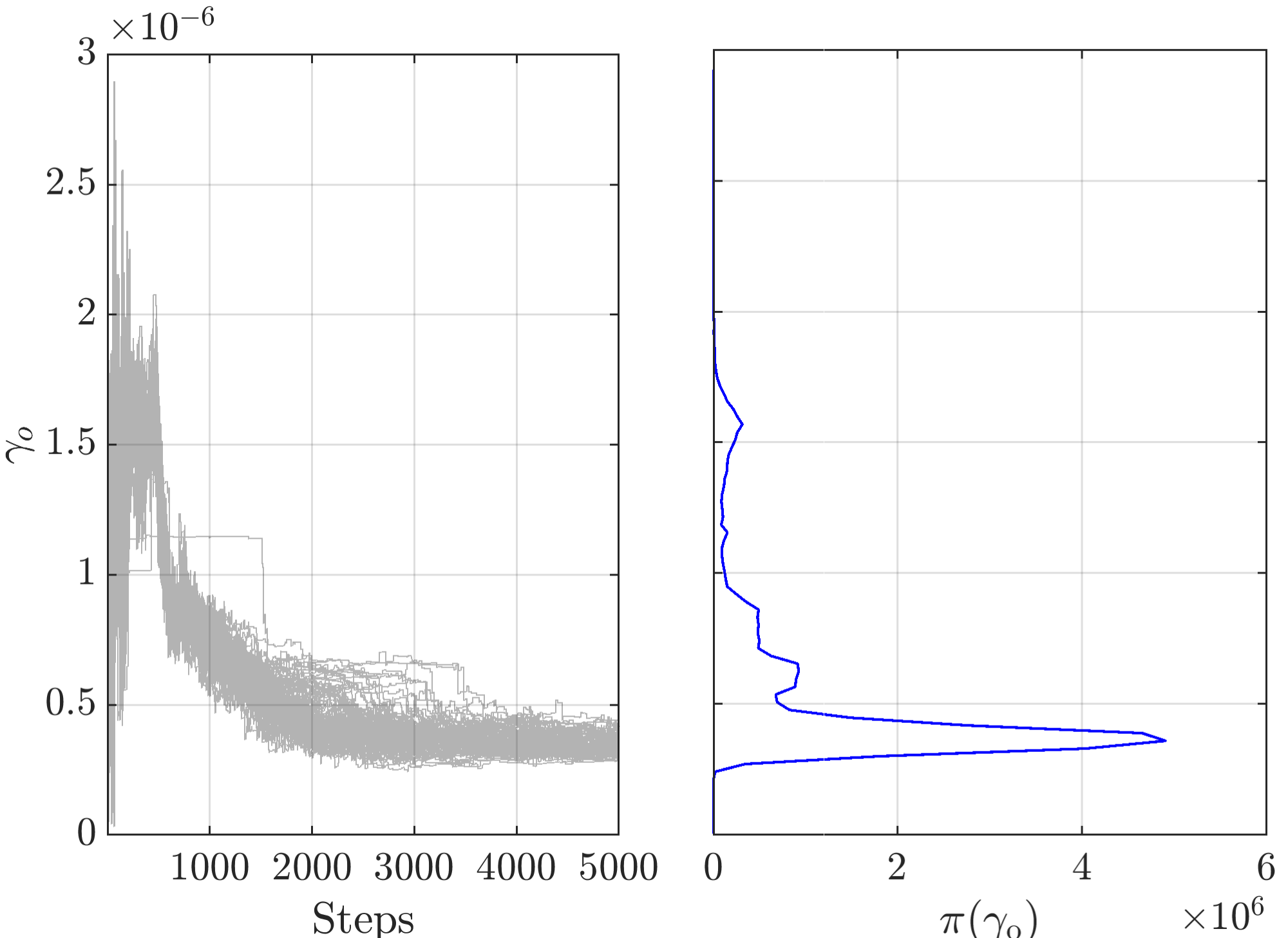}}
    \caption{Trace plots of the parameters}
    \label{fig:trace_1}
\end{figure}
\begin{figure}[h!]
    \centering
    \subfigure[]{\includegraphics[width=0.49\linewidth]{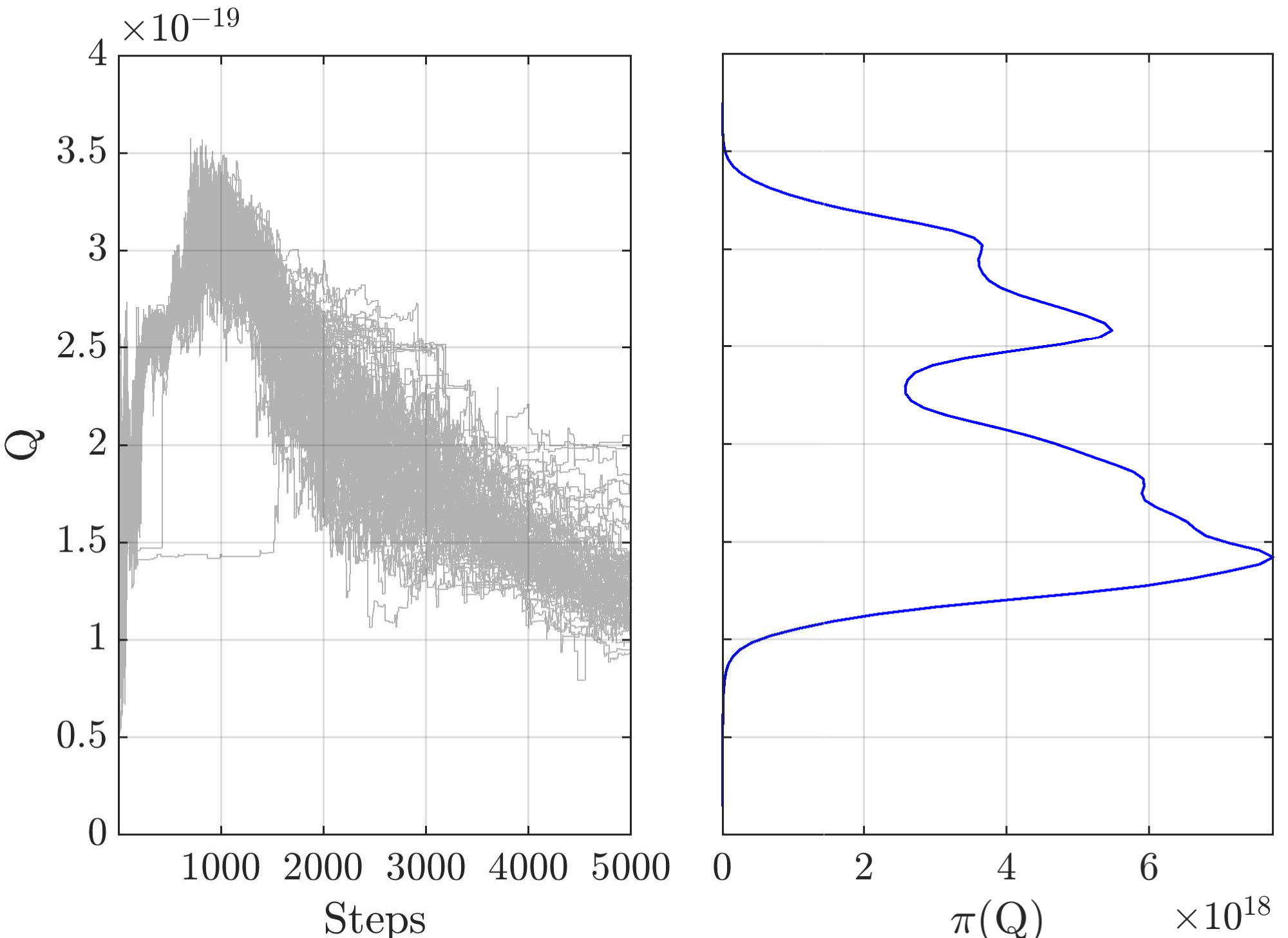}}
    \subfigure[]{\includegraphics[width=0.49\linewidth]{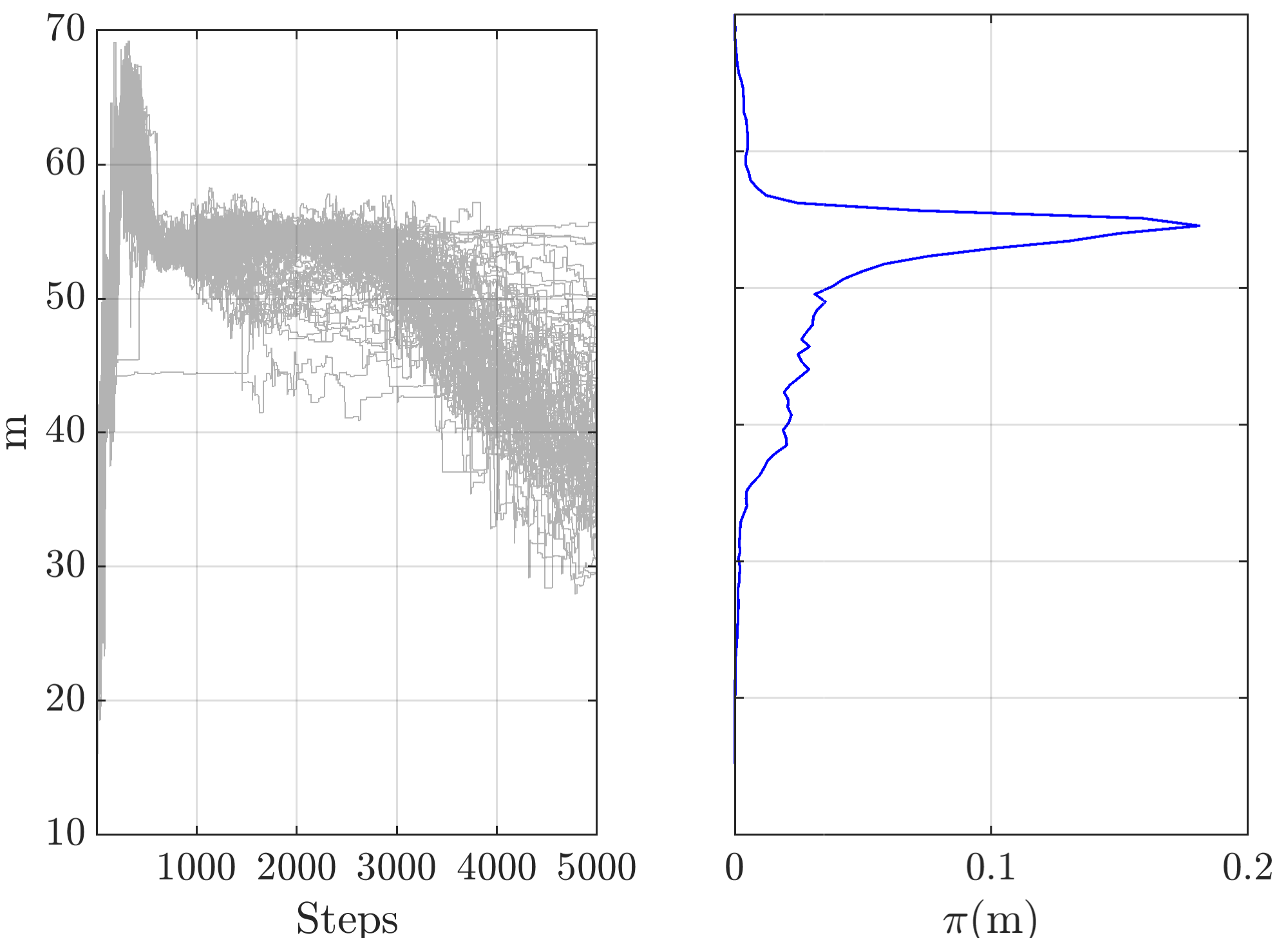}}
    \caption{Trace plots of the parameters}
    \label{fig:trace_2}
\end{figure}
The convergence of the mean of the parameters is shown in Fig.~\ref{fig:mean_1}. The mean convergence plots show the same trend as the trace plots.
\begin{figure}[h!] \ContinuedFloat
    \centering
    \subfigure[]{\includegraphics[width=0.49\linewidth]{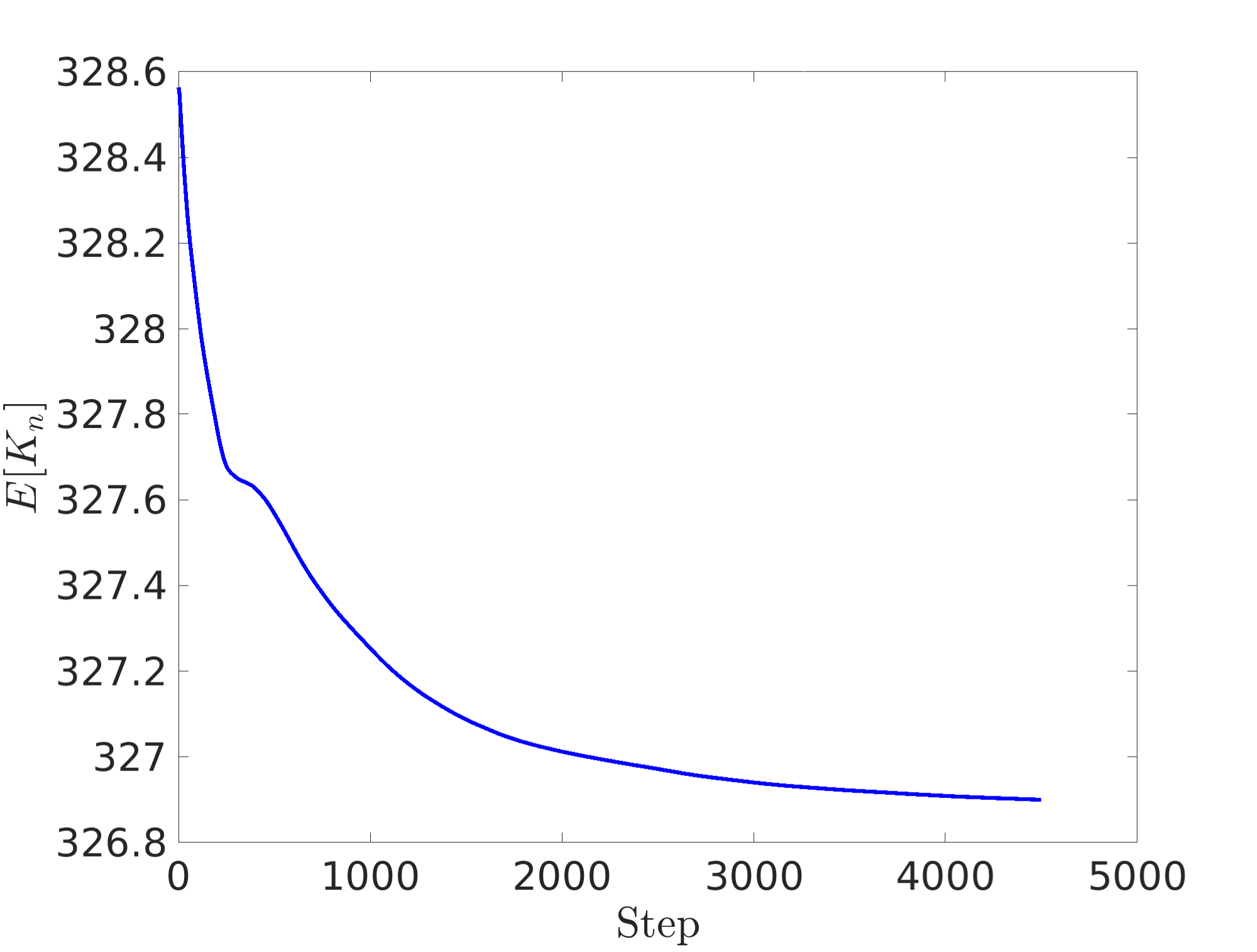}}
    \subfigure[]{\includegraphics[width=0.49\linewidth]{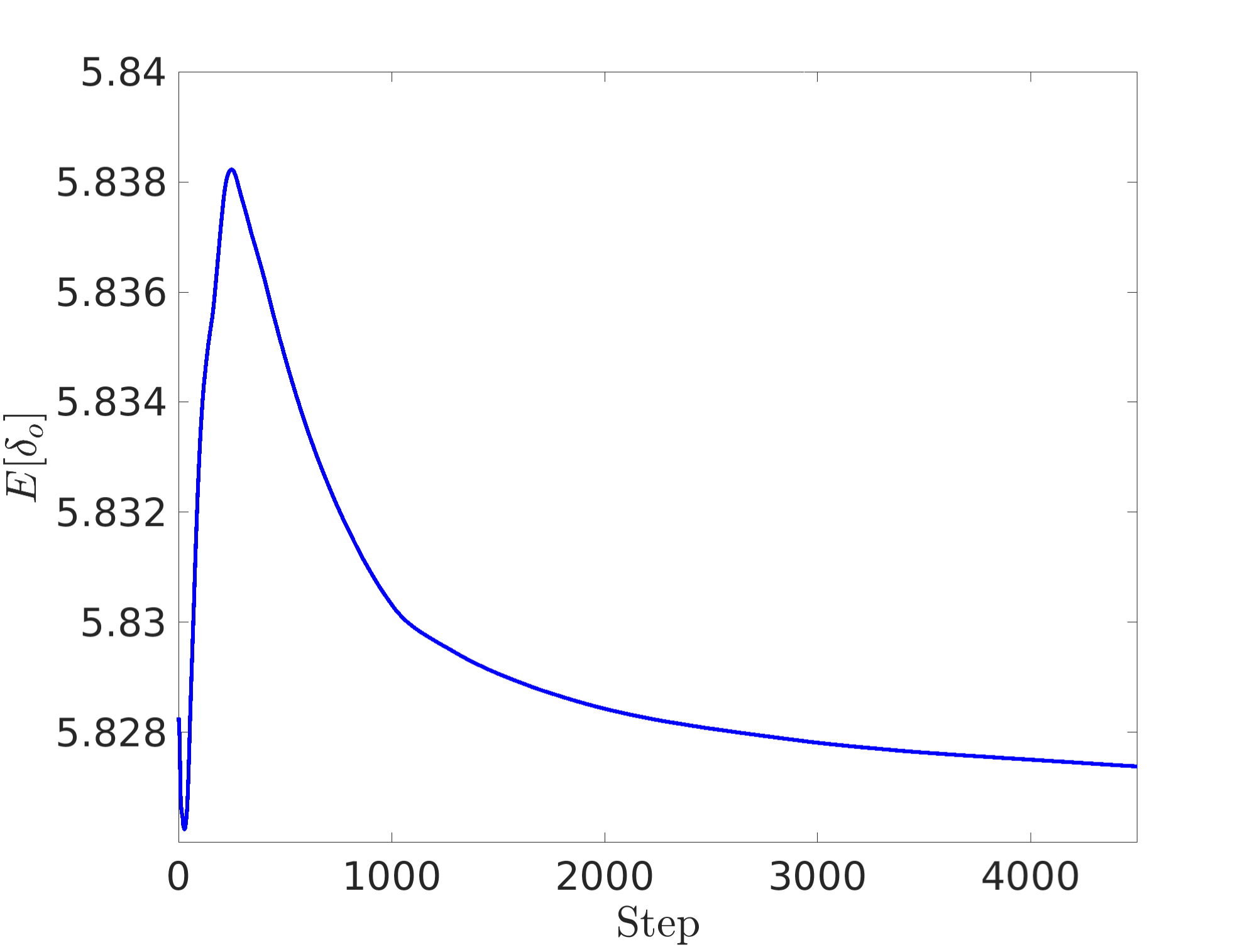}}
    \subfigure[]{\includegraphics[width=0.49\linewidth]{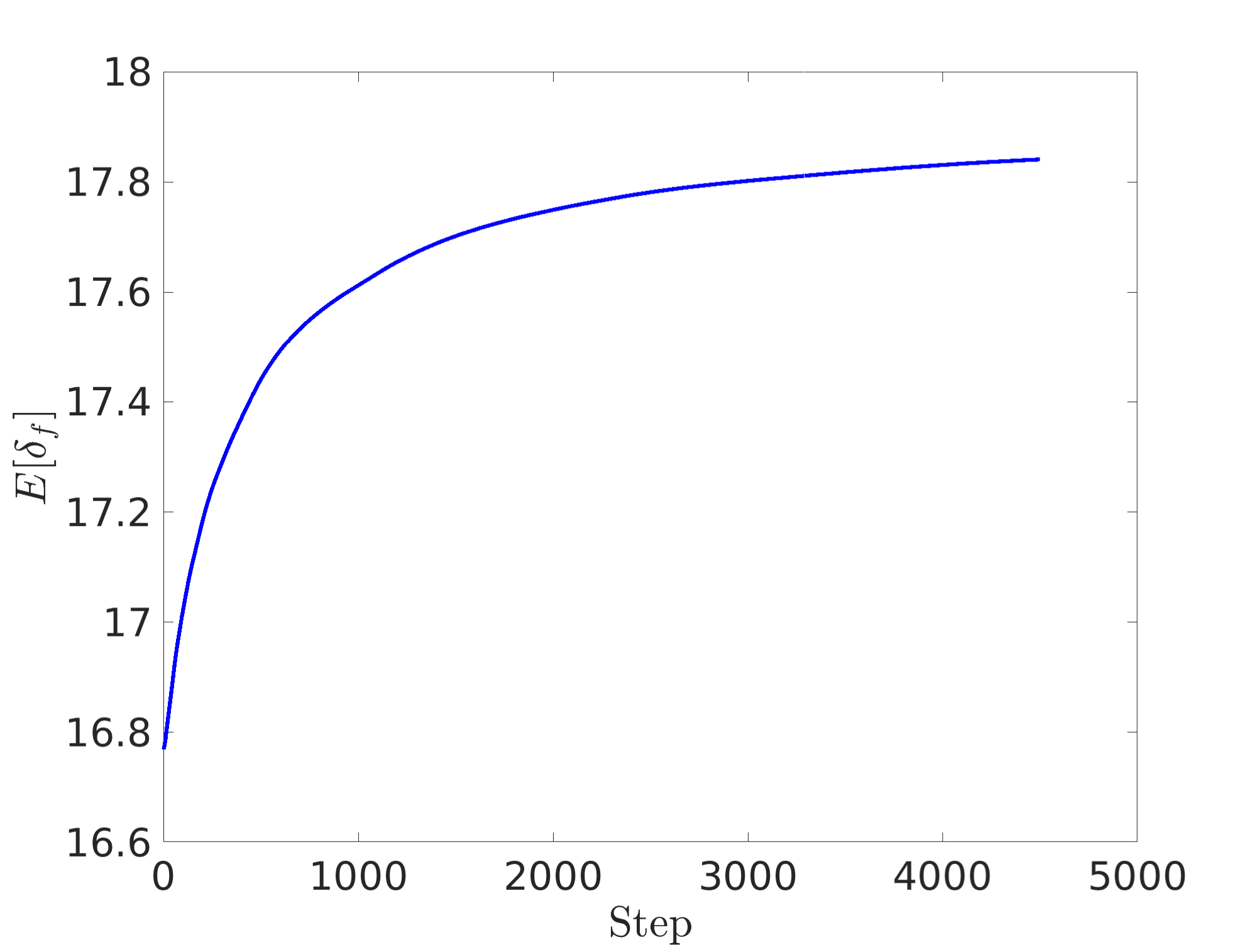}}
    \subfigure[]{\includegraphics[width=0.49\linewidth]{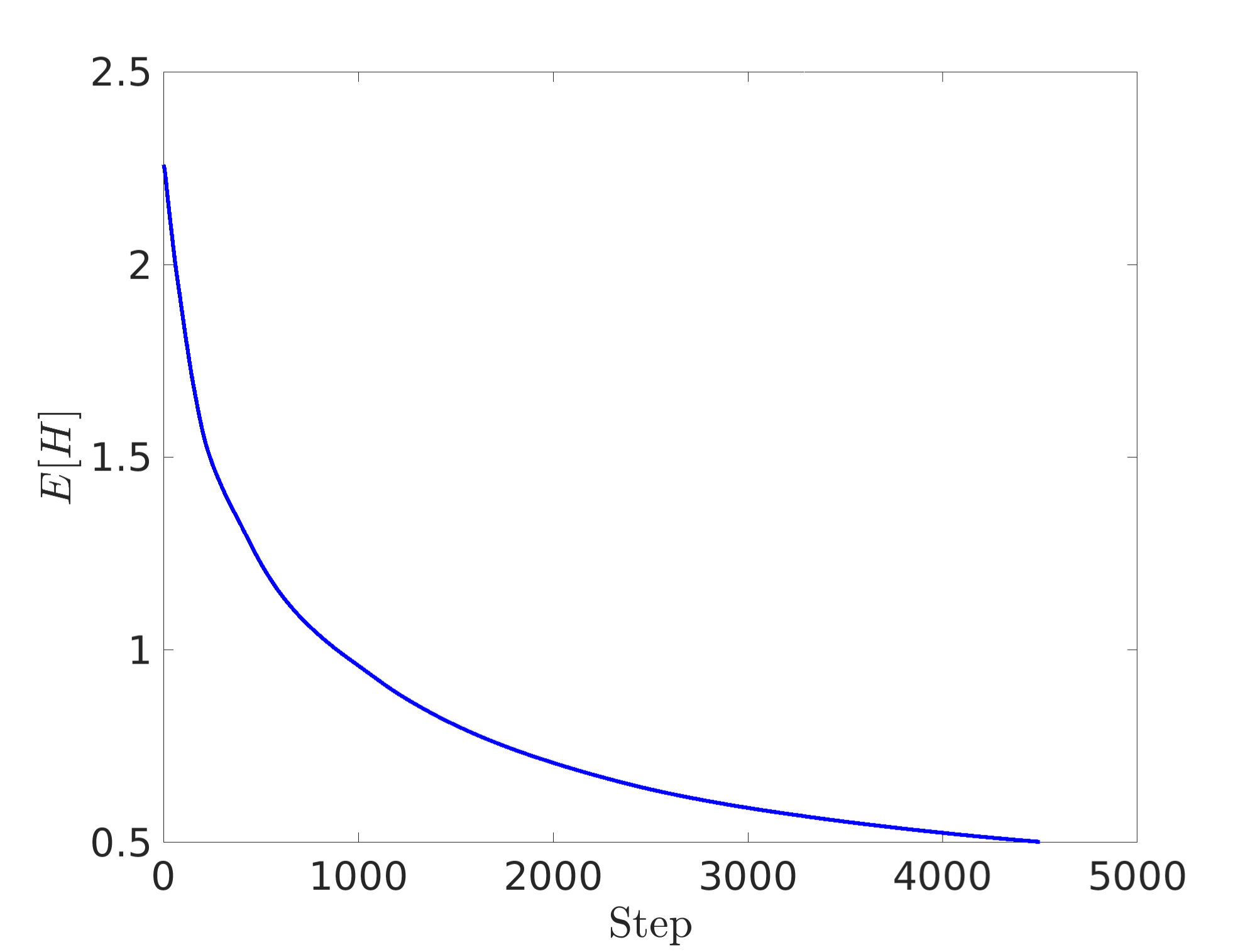}}
    \subfigure[]{\includegraphics[width=0.49\linewidth]{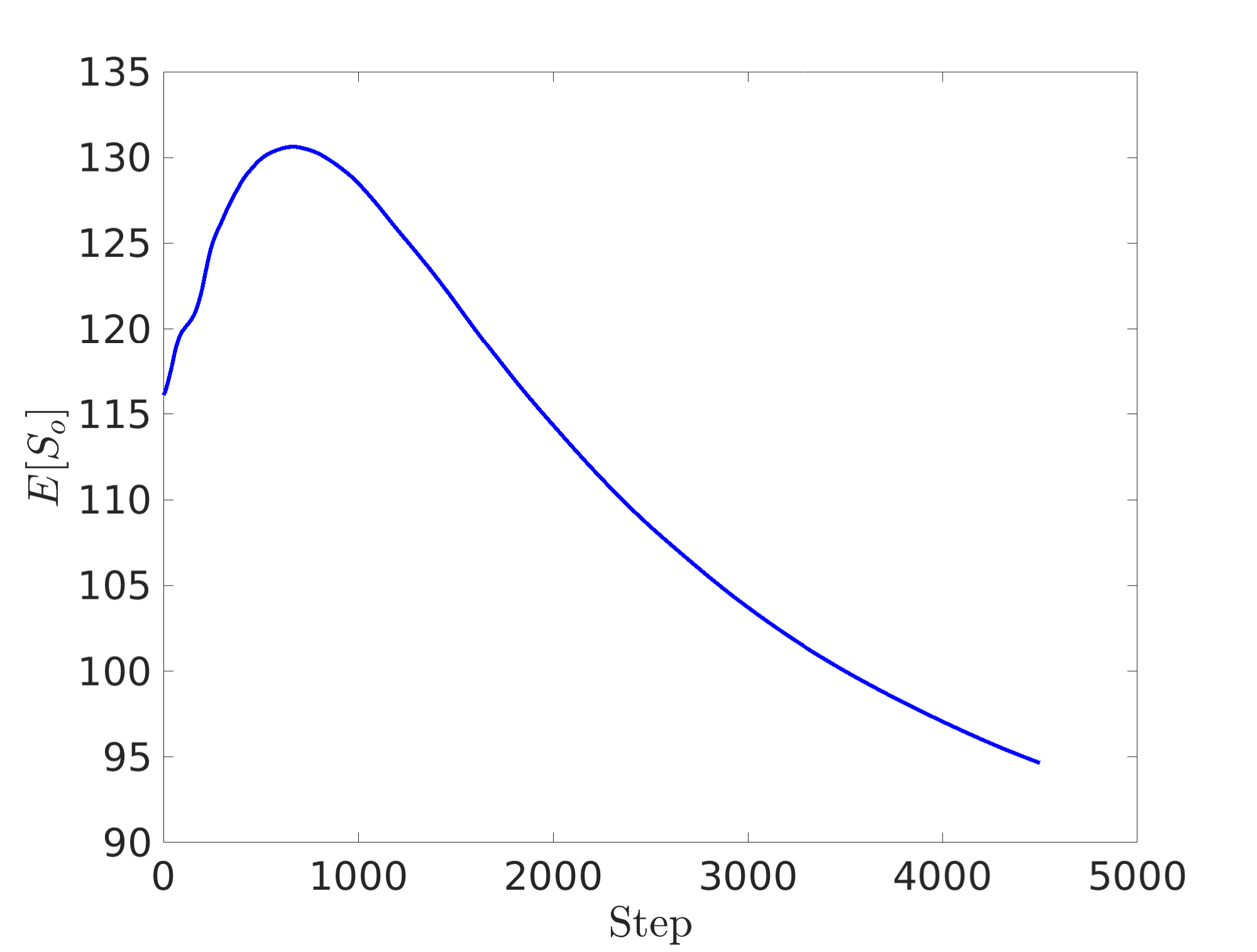}}
    \subfigure[]{\includegraphics[width=0.49\linewidth]{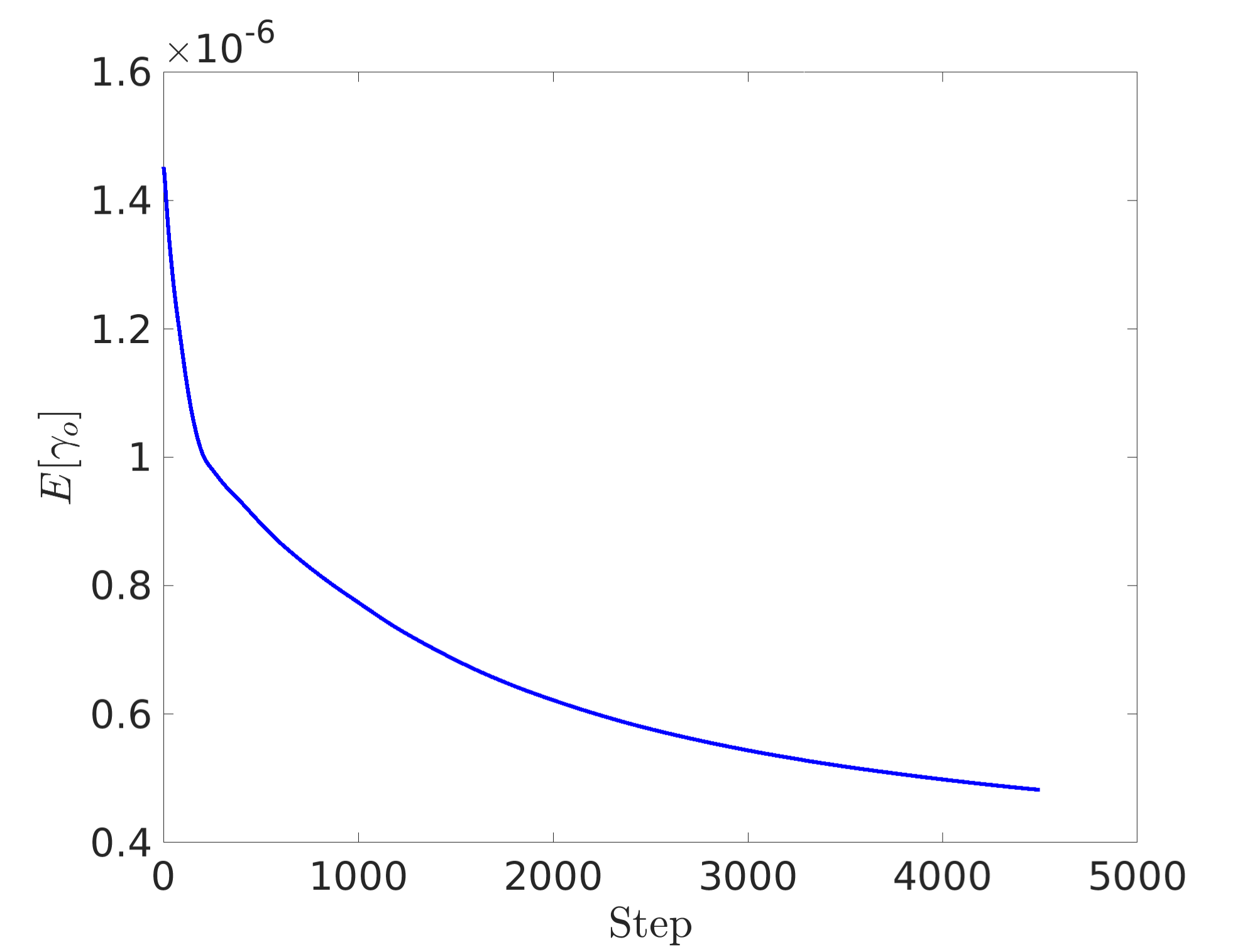}}
    \caption{Convergence of the means of the prameters}
    \label{fig:mean_1}
\end{figure}
\begin{figure}[h!]
    \centering
    \subfigure[]{\includegraphics[width=0.49\linewidth]{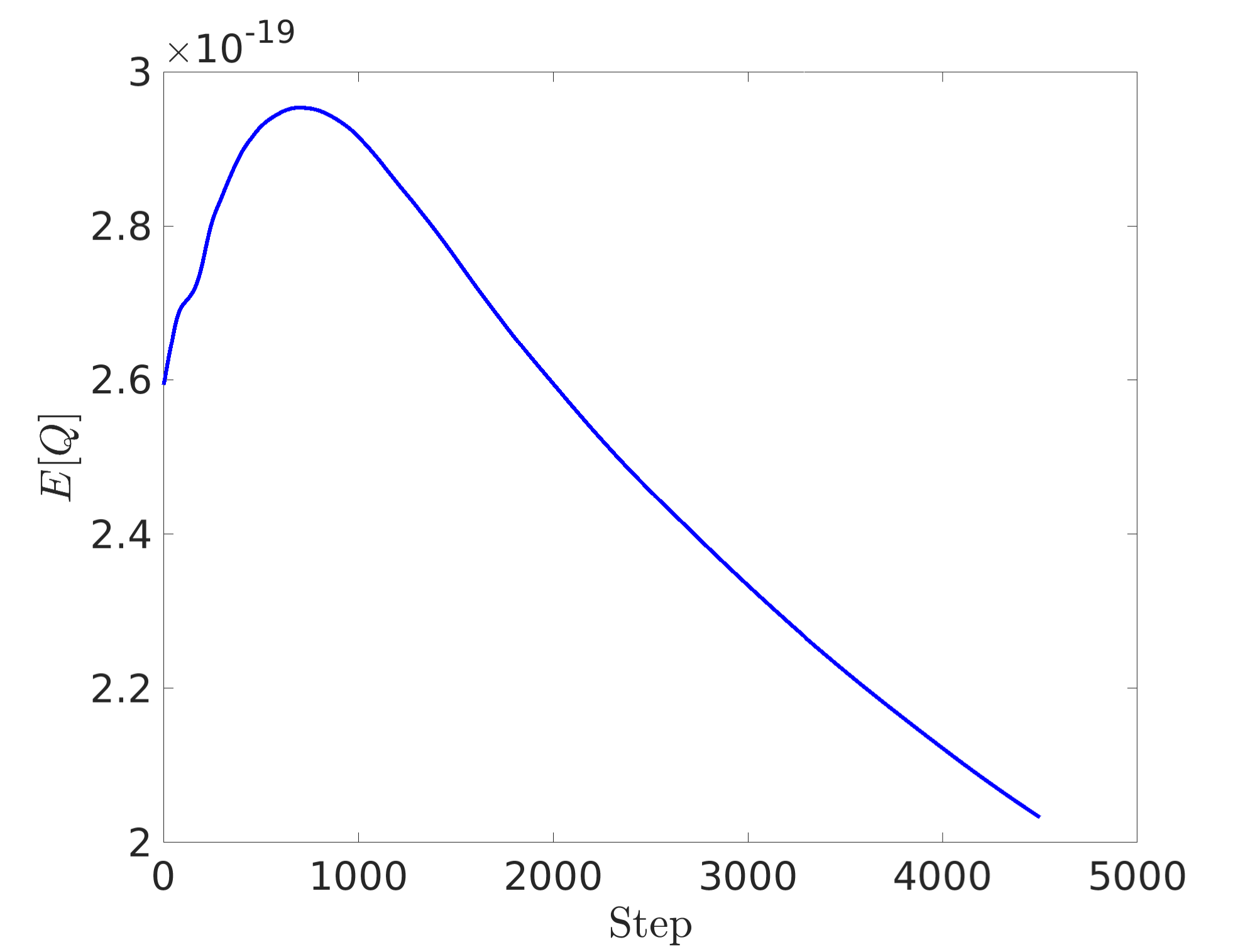}}
    \subfigure[]{\includegraphics[width=0.49\linewidth]{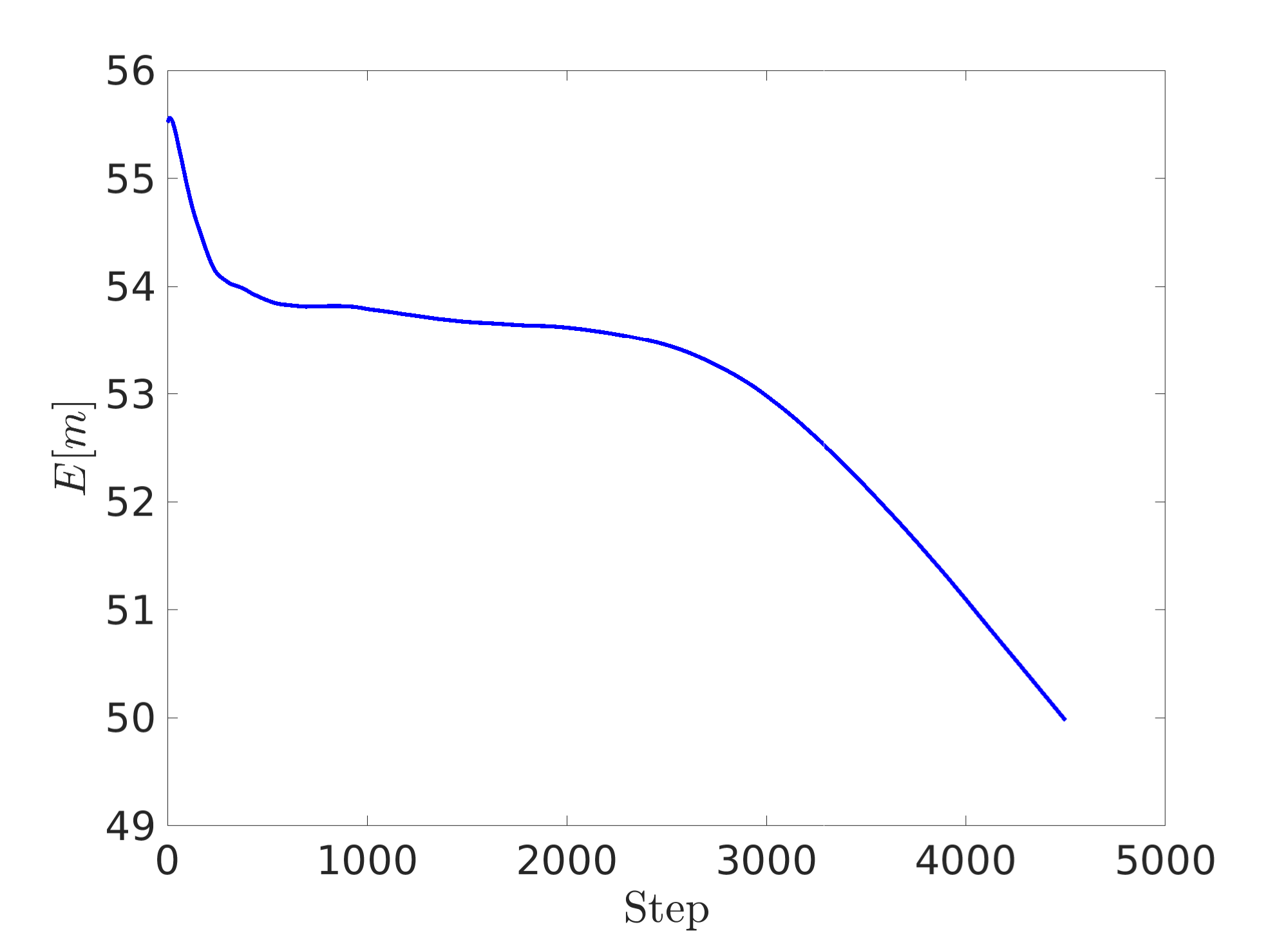}}
    \caption{Convergence of the means of the prameters}
    \label{fig:mean_2}
\end{figure}

\begin{figure}[h!]
    \centering
    \subfigure[Prior samples of the Bayesian Calibration]{\label{fig:prior_samp_5}\includegraphics[width=\linewidth]{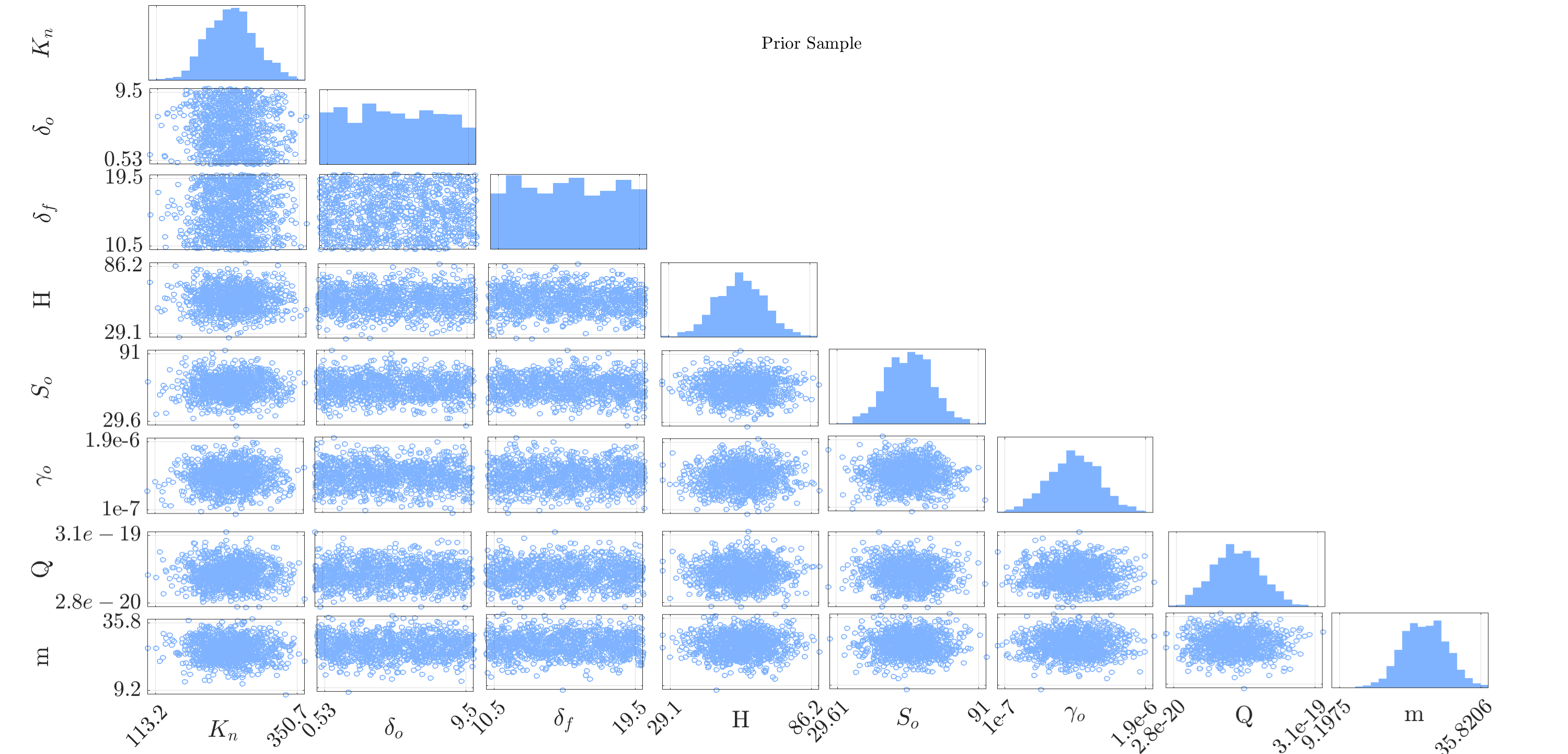}}
    \subfigure[Posterior samples of the Bayesian Calibration]{\label{fig:post_samp_5}\includegraphics[width=\linewidth]{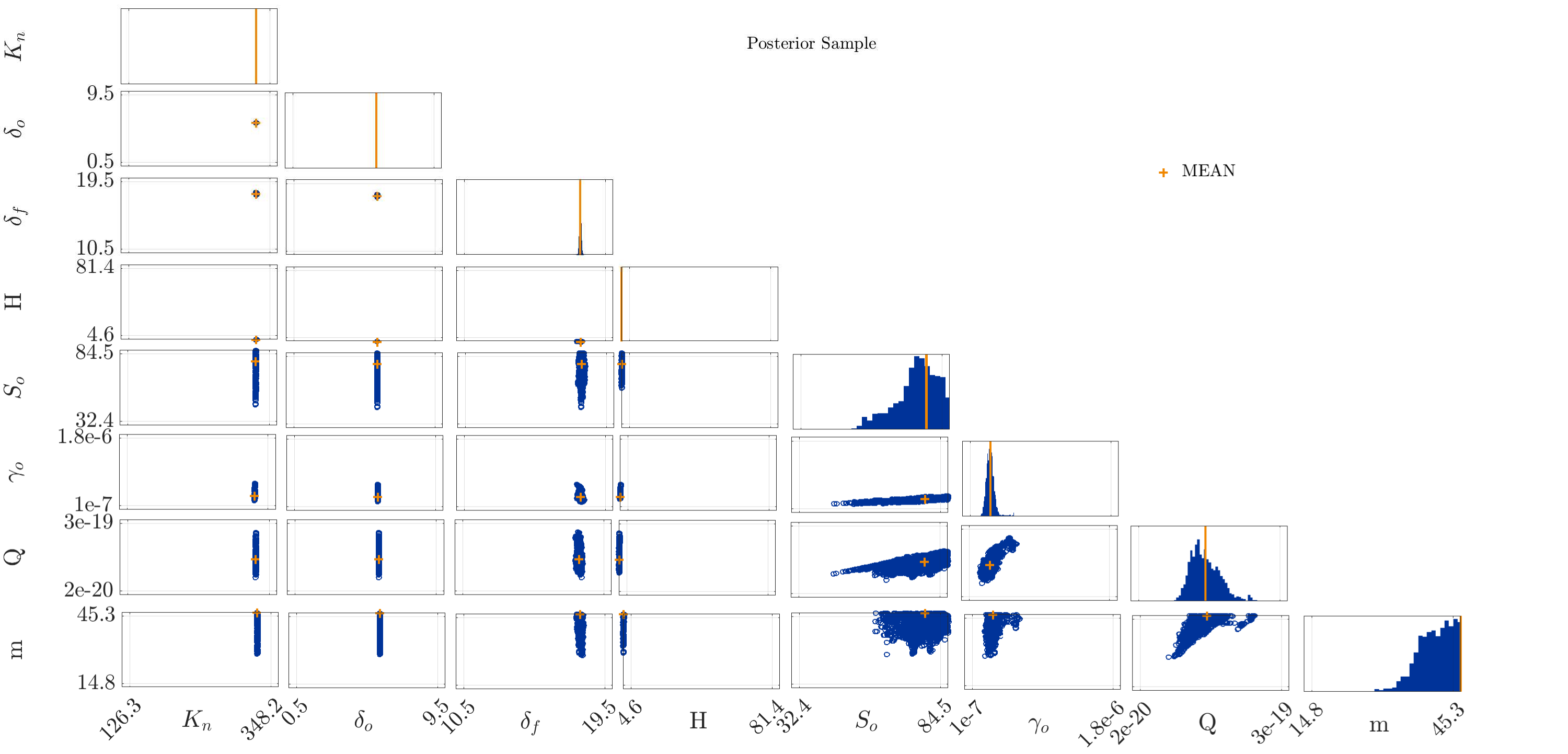}}
    \caption{Samples drawn from prior and posterior distributions during calibration.}
    \label{fig:samp_calib}
\end{figure}

A violin plot is a statistical graph showing the probability density of a quantity. Violin plots depicting the posterior predictive distribution of the load-displacement curve are presented in Figure \ref{fig:post_pred}. The effect of the parameter uncertainty is seen from the violin plots. It is evident that the parameter uncertainties are not sufficient to capture the experimental response. This emphasizes the need for a discrepancy term.
\begin{figure}[h!] \ContinuedFloat
    \centering
    \subfigure[5 mm/min]{\includegraphics[width=0.75\linewidth]{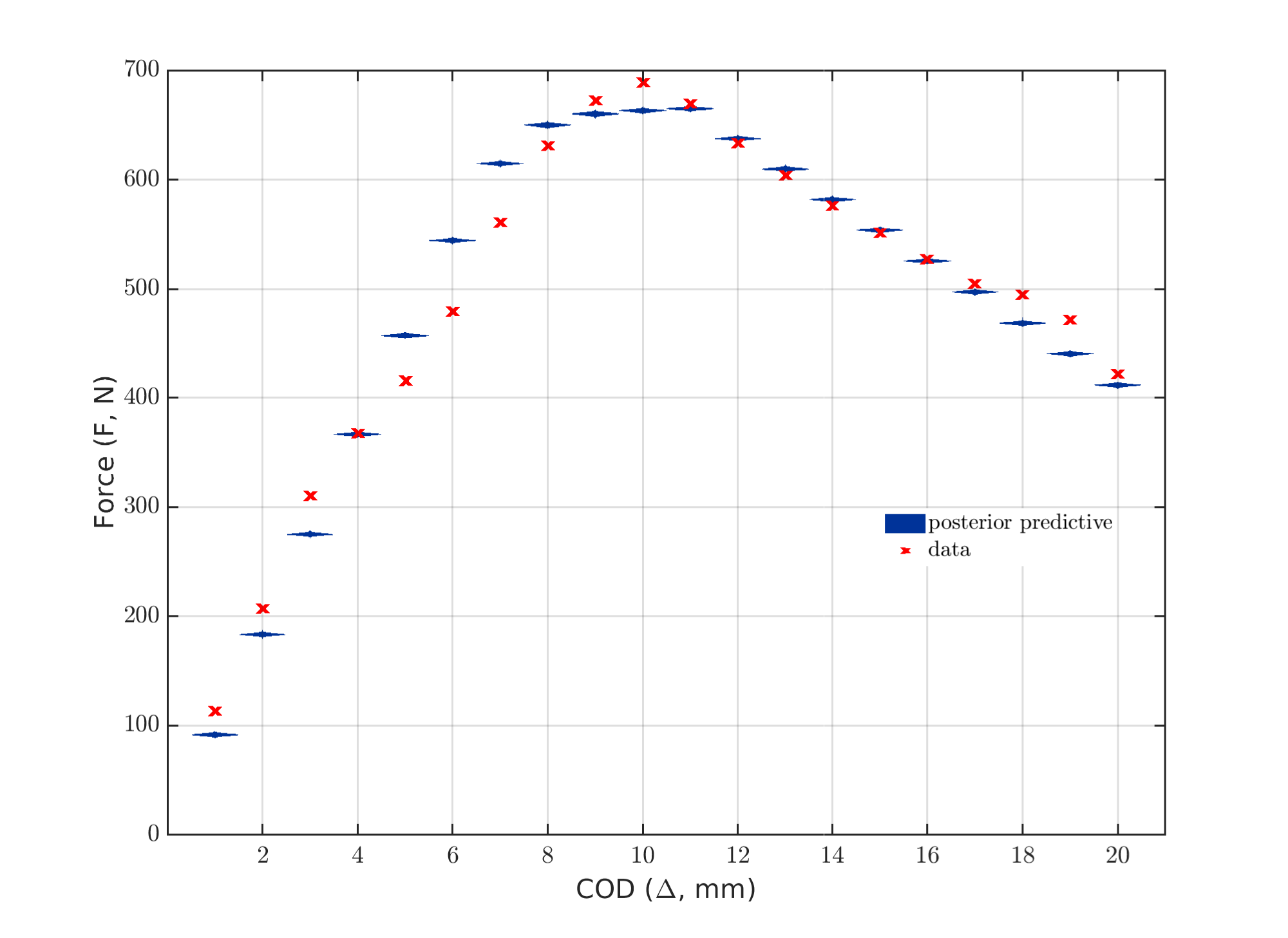}}
    \subfigure[50 mm/min]{\includegraphics[width=0.75\linewidth]{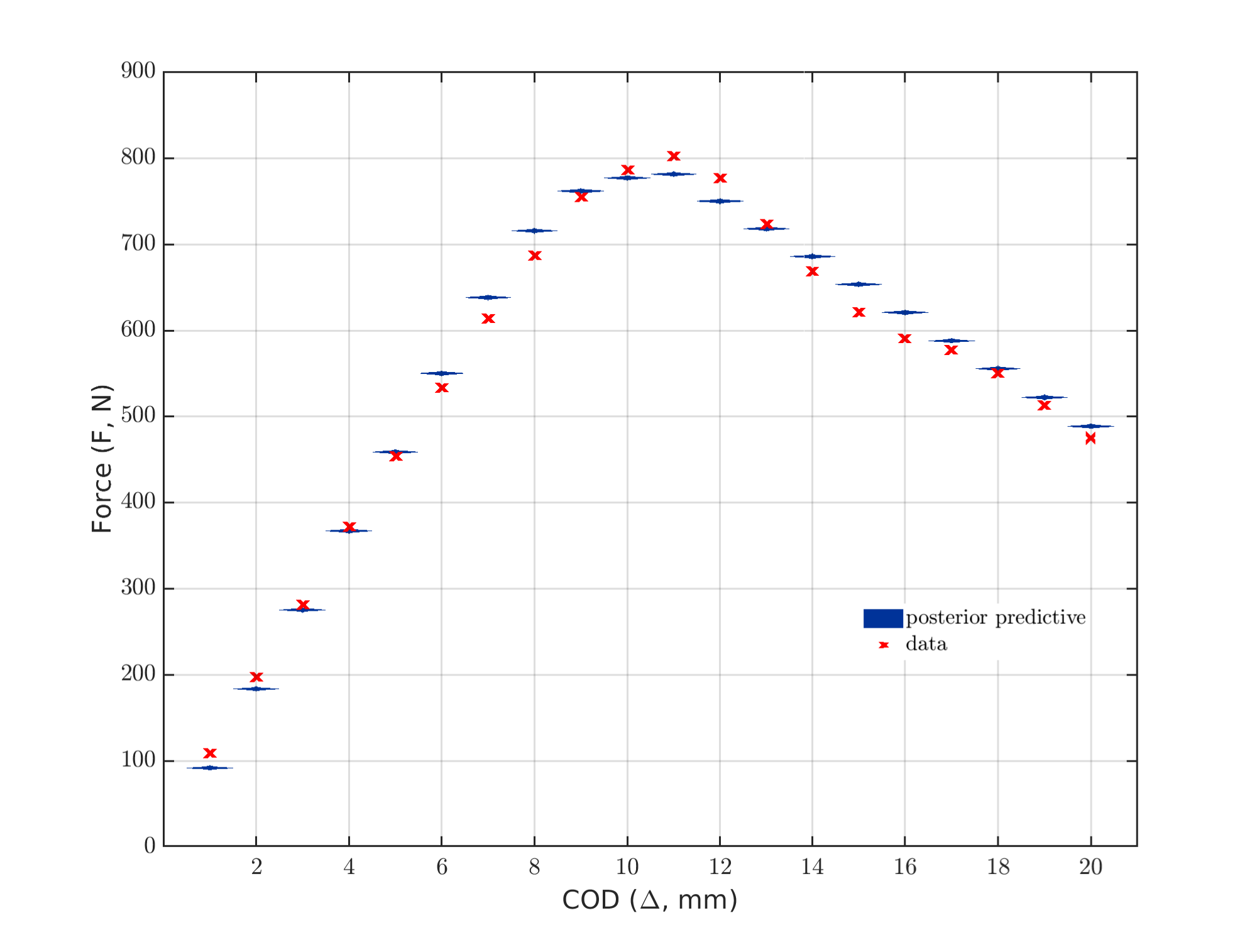}}
    \caption{Posterior predictive distribution after Bayesian Calibration}
    \label{fig:post_pred}
\end{figure} 
\begin{figure}[h!] 
    \centering
    \subfigure[500 mm/min]{\includegraphics[width=0.75\linewidth]{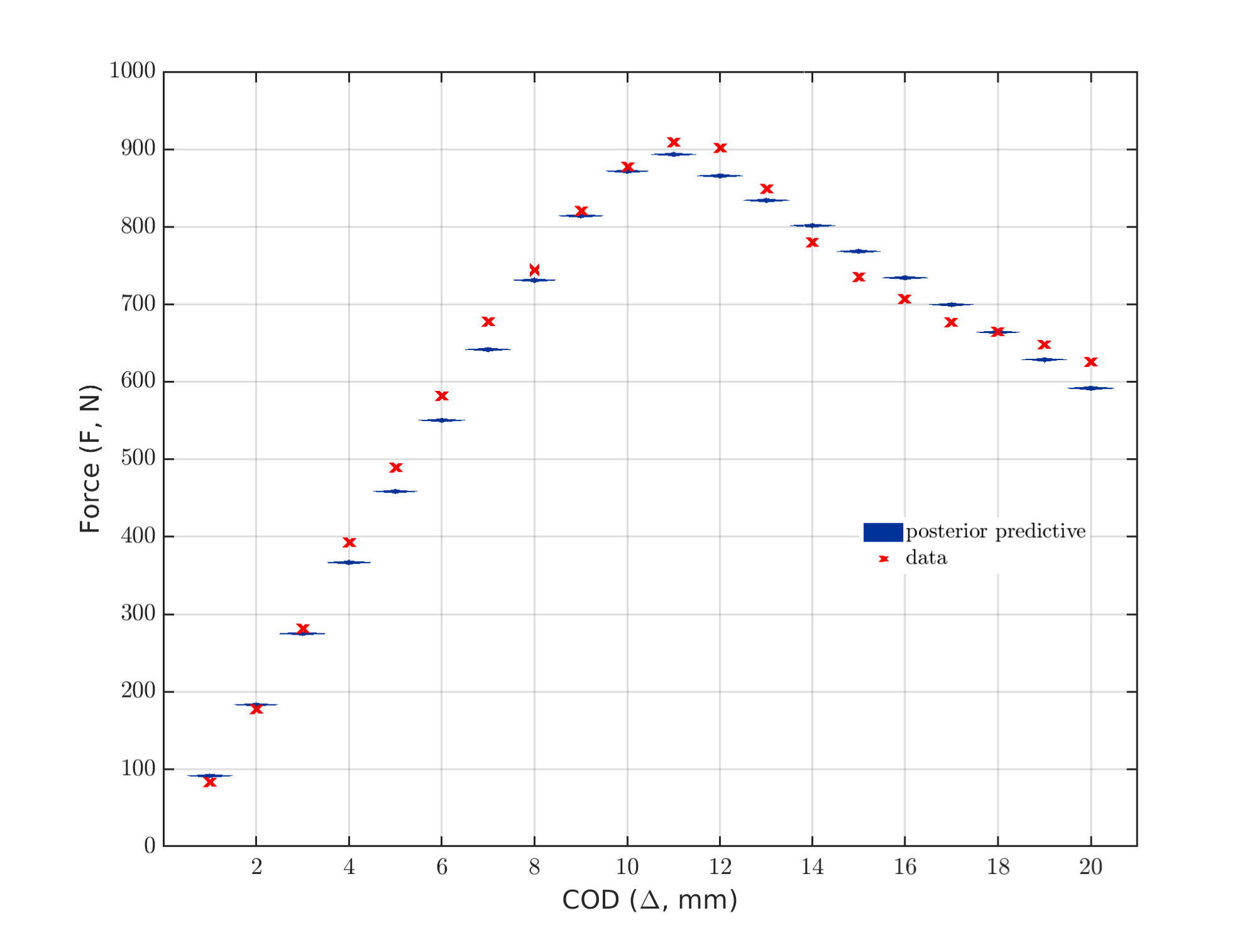}}
    \caption{Posterior predictive distribution after Bayesian Calibration}
    \label{fig:post_pred1}
\end{figure}
The samples drawn from the prior and posterior distribution of parameters are presented through a scatterplot in Figure \ref{fig:samp_calib}. The variance of these samples is a measure of uncertainty in the parameters of the model.

\section{Error convergence}\label{app:conv}
\begin{figure}[htpb]
    \centering
    \includegraphics[width = 0.8\linewidth]{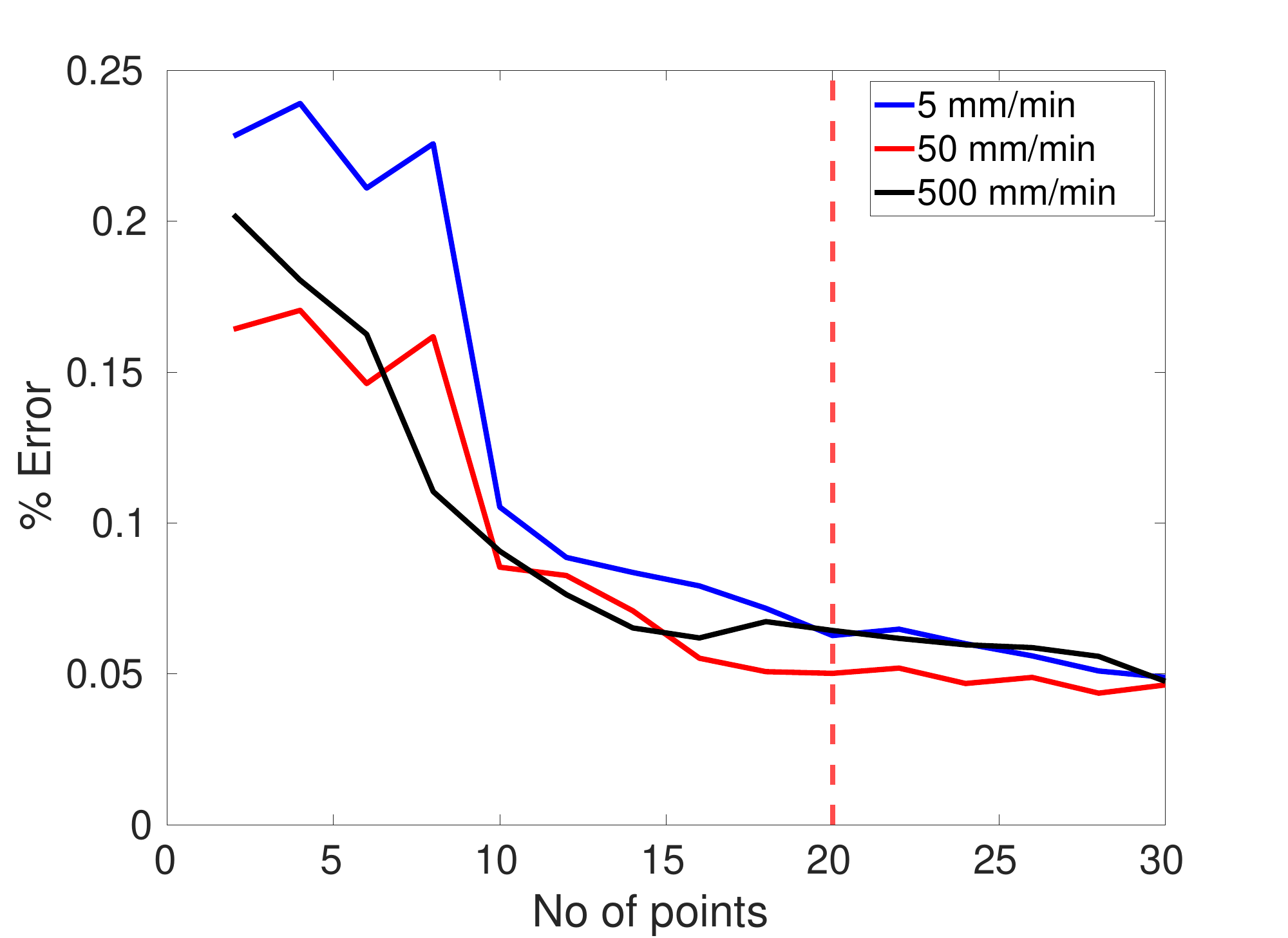}
    \caption{Convergence of error with the number of training points in discrepancy.}
    \label{fig:error_conv}
\end{figure}

A convergence study for the error is performed to select the optimal number of points needed to learn the discrepancy function. The plot of the convergence of the percentage error is shown in Fig.~\ref{fig:error_conv}.

\end{document}